\newif\ifTechRep
\TechReptrue
\newif\ifShowComments
\ShowCommentstrue	

\documentclass[sigconf, nonacm]{acmart}
\usepackage[font=small, labelformat=simple]{subcaption}
\usepackage[font=small, labelformat=simple]{caption}
\usepackage[linesnumbered, vlined, boxed, ruled]{algorithm2e}
\usepackage{enumitem}

\usepackage{hyperref}
\usepackage{color}
\usepackage{multicol}
\usepackage{multirow}

\usepackage{array}
\usepackage{color, colortbl}
\usepackage{bbm}
\usepackage{booktabs}
\usepackage{balance}
\usepackage{rotating}
\usepackage{alltt}
\usepackage{hyperref}
\usepackage{pdfrender}
\usepackage{stmaryrd}
\usepackage{anyfontsize}
\usepackage{amsthm}
\usepackage{soul}
\usepackage{xcolor}
\usepackage{soul}
\usepackage{xspace}
\usepackage{makecell}
\usepackage{anyfontsize}
\usepackage{silence,etoolbox}
\usepackage{subcaption}
\SetCommentSty{mycommfont}

\newcommand{\turnrow}{\footnotesize\rotatebox{90}}

\makeatletter
\patchcmd{\wrong@fontshape}{\@gobbletwo}{}{}{}
\makeatother
\newcommand\vldbdoi{XX.XX/XXX.XX}
\newcommand\vldbpages{XXX-XXX}
\newcommand\vldbvolume{14}
\newcommand\vldbissue{1}
\newcommand\vldbyear{2020}
\newcommand\vldbauthors{\authors}
\newcommand\vldbtitle{\shorttitle} 

\newcommand\vldbpagestyle{plain} 

\sethlcolor{white}
\setlist{nolistsep,leftmargin=*}
\definecolor{vlightgray}{gray}{0.85}
\let\paragraph\relax

\SetKwRepeat{Do}{do}{while}

\newtheorem{theorem}{Theorem}
\newtheorem{definition}[theorem]{Definition}
\newtheorem{problem}[theorem]{Problem}

\newtheorem{example}[theorem]{Example}

\newcommand{\paragraph}[1]{\noindent\textbf{#1}}

\newcommand{\ignore}[1]{{}}

\newcommand{\algoexp}[1]{{\noindent \textbf{#1}}}
\newcommand{\system}{\textsc{DataExposer}\xspace}
\newcommand{\systemgr}{\textsc{DataExposer}\textsubscript{GRD}\xspace}
\newcommand{\systemgp}{\textsc{DataExposer}\textsubscript{GT}\xspace}

\ifShowComments
\newcommand{\sg}[1]{[\textcolor{magenta}{SG: #1}]}
\newcommand{\rl}[1]{[\textcolor{red}{RL: #1}]}
\newcommand{\af}[1]{[\textcolor{blue}{AF: #1}]}

\newcommand{\am}[1]{\textcolor{teal}{[AM: #1]}}
\else
\newcommand{\sg}[1]{}
\newcommand{\rl}[1]{}
\newcommand{\af}[1]{}
\newcommand{\am}[1]{}
\fi

\newcommand{\pl}[1]{#1}

\def\Dom{\texttt{Dom}}
\def\D{\mathcal{D}}
\def\E{{\mathcal{X}^{*}}}
\def\DDom{\mathbf{Dom}}

\def\P{\mathcal{X}}
\def\T{T}
\DeclareMathOperator{\argmax}{arg\,max}

\newcommand{\domain}{\textsc{Domain}\xspace}
\newcommand{\correlation}{\textsc{Indep}\xspace}

\newcommand{\outlier}{\textsc{Outlier}\xspace}
\newcommand{\selectivity}{\textsc{Selectivity}\xspace}
\newcommand{\missing}{\textsc{Missing}\xspace}
\newcommand{\pplfail}{$\mathit{People_{fail}}$\xspace}
\newcommand{\pplpass}{$\mathit{People_\mathit{pass}}$\xspace}

\begin{document}
\title[\system: Exposing Disconnect between Data and Systems]{\system: Exposing Disconnect between Data and Systems}

\author{Sainyam Galhotra}
\affiliation{
  \institution{University of Massachusetts Amherst}
}
\email{sainyam@cs.umass.edu}

\author{Anna Fariha}
\affiliation{
  \institution{University of Massachusetts Amherst}
}
\email{afariha@cs.umass.edu}

\author{Raoni Louren\c{c}o}
\affiliation{
  \institution{New York University}
}
\email{raoni@nyu.edu}

\author{Juliana Freire}
\affiliation{
  \institution{New York University}
}
\email{juliana.freire@nyu.edu}

\author{Alexandra Meliou}
\affiliation{
  \institution{University of Massachusetts Amherst}
}
\email{ameli@cs.umass.edu}

\author{Divesh Srivastava}
\affiliation{
  \institution{AT\&T Chief Data Office}
}
\email{divesh@att.com}

\begin{abstract}
%

\looseness-1 As data is a central component of many modern systems, the cause
of a system malfunction may reside in the data, and, specifically, particular
properties of the data. For example, a health-monitoring system that is
designed under the assumption that weight is reported in imperial units (lbs)
will malfunction when encountering weight reported in metric units (kilograms).
Similar to software debugging, which aims to find bugs in the mechanism (source
code or runtime conditions), our goal is to debug the data to identify
potential sources of \emph{disconnect} between the assumptions about the data
and the systems that operate on that data. Specifically, we seek which
\emph{properties} of the data cause a data-driven system to malfunction. We
propose \system, a framework to identify data properties, called
\emph{profiles}, that are the root causes of performance degradation or failure
of a system that operates on the data. Such identification is necessary to
repair the system and resolve the disconnect between data and system. Our
technique is based on \emph{causal reasoning} through \emph{interventions}:
when a system malfunctions for a dataset, \system alters the data profiles and
observes changes in the system's behavior due to the alteration. Unlike
statistical observational analysis that reports mere correlations, \system
reports causally verified root causes---in terms of data profiles---of the
system malfunction. We empirically evaluate \system on three real-world and
several synthetic data-driven systems that fail on datasets due to a diverse
set of reasons. In all cases, \system identifies the root causes precisely
while requiring orders of magnitude fewer interventions than prior techniques.

\end{abstract}

\maketitle

\pagestyle{\vldbpagestyle}
 \begingroup\small\noindent\raggedright\textbf{PVLDB Reference Format:}\\
 \vldbauthors. \vldbtitle. PVLDB, \vldbvolume(\vldbissue): \vldbpages, \vldbyear.\\
 \href{https://doi.org/\vldbdoi}{doi:\vldbdoi}
 \endgroup
 \begingroup
 \renewcommand\thefootnote{}\footnote{\noindent
 This work is licensed under the Creative Commons BY-NC-ND 4.0 International
 License. Visit \url{https://creativecommons.org/licenses/by-nc-nd/4.0/} to view
 a copy of this license. For any use beyond those covered by this license,
 obtain permission by emailing \href{mailto:info@vldb.org}{info@vldb.org}.
 Copyright is held by the owner/author(s). Publication rights licensed to the
 VLDB Endowment. \\
 \raggedright Proceedings of the VLDB Endowment, Vol. \vldbvolume, No. \vldbissue\ %
 ISSN 2150-8097. \\
 \href{https://doi.org/\vldbdoi}{doi:\vldbdoi} \\
 }\addtocounter{footnote}{-1}\endgroup


\section{Introduction}\label{sec:introduction}
%

\looseness-1 Traditional software debugging aims to identify errors and bugs in
the mechanism---such as source code, configuration files, and runtime
conditions---that may cause a system to malfunction~\cite{hailpern2002software,
liblit2005scalable, aid}. However, in modern systems, data has become a central
component that itself can cause a system to fail. Data-driven systems comprise
complex pipelines that rely on data to solve a target task. Prior work
addressed the problem of debugging machine-learning
models~\cite{cadamuro2016debugging} and finding root causes of failures in
computational pipelines~\cite{bugdoc}, where certain values of the pipeline
parameters---such as a specific model and/or a specific dataset---cause the
pipeline failure. However, just knowing that a pipeline fails for a certain
dataset is not enough; naturally, we ask: what \emph{properties} of a dataset
caused the failure?

\looseness-1 Two common reasons for malfunctions in data-driven systems are:
(1)~incorrect data, and (2)~\emph{disconnect} between the assumptions about the
data and the design of the system that operates on the data. Such disconnects
may happen when the system is not robust, i.e., it makes strict assumptions
about metadata (e.g., data format, domains, ranges, and distributions), and
when new data drifts from the data over which the system was tested on before
deployment~\cite{rawal2020can} (e.g., when a system expects a data stream to
have a weekly frequency, but the data provider suddenly switches to daily data).

Therefore, in light of a failure, one should investigate potential issues in
the data. Some specific examples of commonly observed system malfunctions
caused by data include: (1)~decline of a machine-learned model's accuracy (due
to out-of-distribution data), (2)~unfairness in model predictions (due to
imbalanced training data), (3)~excessive processing time (due to a system's
failure to scale to large data), and (4)~system crash (due to invalid input
combination in the data tuples beyond what the system was designed to handle).
These examples indicate a common problem: \emph{disconnect} or \emph{mismatch}
between the data and the system design. Once the mismatch is identified, then
possible fixes could be either to repair the data to suit the system design, or
to adjust the system design (e.g., modify source code) to accommodate data with
different properties.

\looseness-1
A na\"{\i}ve approach to deal with potential issues in the data is to identify
outliers: report tuples as potentially problematic based on how atypical they
are with respect to the rest of the tuples in the dataset. However, without
verifying whether the outliers actually cause unexpected outcomes, we can never
be certain about the actual root causes. As pointed out in prior
work~\cite{DBLP:conf/oopsla/BarowyGB14}: \emph{``With respect to a computation,
whether an error is an outlier in the program's input distribution is not
necessarily relevant. Rather, potential errors can be spotted by their
\underline{effect} on a program's output distribution.''} To motivate our work,
we start with an example taken from a real-world incident, where Amazon's
delivery service was found to be racist~\cite{amazonunfair}.

\vspace{-1mm}
\begin{example}[Biased Classifier]\label{ex:one}
An e-commerce company wants to build an automated system that suggests who
should get discounts. To this end, they collect information from the customers'
purchases over one year and build a dataset over the attributes \texttt{name},
\texttt{gender}, \texttt{age}, \texttt{race}, \texttt{zip\_code},
\texttt{phone}, \texttt{products\_purchased}, etc.
Anita, a data scientist, is then asked to develop a machine learning (ML)
pipeline over this dataset to predict whether a customer will spend over a
certain amount, and, subsequently, should be offered discounts. Within this
pipeline, Anita decides to use a logistic regression classifier
for prediction and implements it using an off-the-shelf ML library.
%
To avoid discrimination over any group and to ensure that the classifier
trained on this dataset is fair, Anita decides to drop the sensitive
attributes---\texttt{race} and \texttt{gender}---during the pre-processing step
of the ML pipeline, before feeding it to the classifier. However, despite this
effort, the trained classifier turns out to be highly biased against African
American people and women. After a close investigation, Anita discovers that:
(1)~In the training data, \texttt{race} is highly correlated with
\texttt{zip\_code}, and (2)~The training dataset is imbalanced: a larger
fraction of the people who purchase expensive products are male. Now she
wonders: if these two properties did not hold in the dataset, would the learned
classifier be fair? Have either (or both) of these properties caused the
observed unfairness?
\end{example}

\looseness-1 Unfortunately, existing tools (e.g.,
CheckCell~\cite{DBLP:conf/oopsla/BarowyGB14}) that blame individual cells
(values) for unexpected outcomes cannot help here, as no single cell in the
training data is responsible for the observed discrimination, rather, global
statistical properties (e.g., correlation) that involve multiple attributes
over the entire data are the actual culprits. Furthermore, Anita only
identified two potential or \emph{correlated} data issues that may or may not
be the actual cause of the unfairness. To distinguish mere correlation from
true causation and to verify if there is indeed a \emph{causal} connection
between the data properties and the observed unfairness, we need to dig deeper.

Example~\ref{ex:one} is one among many incidents in real-world applications
where issues in the data caused systems to malfunction~\cite{amazonhiring,
googleracism}. A recent study of 112 high-severity incidents in Microsoft Azure
services showed that 21\% of the bugs were due to inconsistent assumptions
about data format by different software components or
versions~\cite{DBLP:conf/hotos/LiuLMN19}. The study further found that 83\% of
the data-format bugs were due to inconsistencies between data producers and
data consumers, while 17\% were due to mismatch between interpretations of the
same data by different data consumers. Similar incidents happened due to
misspelling and incorrect date-time
format~\cite{DBLP:conf/cidr/RezigCSSMTOS20}, and issues pertaining to data
fusion where schema assumptions break for a new data
source~\cite{DBLP:journals/pvldb/DongGHHMSZ14, DBLP:conf/sigmod/WangDM15}. We
provide another illustrative example where a system times out when the
distribution of the data, over which the system operates, exhibits significant
skew.

\begin{example}[Process Timeout]\label{ex:two}
\looseness-1 A toll collection software \textsc{EZGo} checks if vehicles
passing through a gate have electronic toll pass installed. If it fails to
detect a toll pass, it uses an external software \textsc{OCR} to extract the
registration number from the vehicle's license plate. \textsc{EZGo} operates in
a batch mode and processes every $1000$ vehicles together by reserving
\textsc{AWS} for one hour, assuming that one hour is sufficient for processing
each batch. However, for some batches, \textsc{EZGo} fails. After a close
investigation, it turns out that the external software \textsc{OCR} uses an
algorithm that is extremely slow for images of black license plates captured in
low illumination. As a result, when a batch contains a large number of such
cases (significantly skewed distribution), \textsc{EZGo} fails.
\end{example}

The aforementioned examples bring forth two key challenges. First, we need to
correctly identify potential causes of unexpected outcomes and generate
\emph{hypotheses} that are expressive enough to capture the candidate root
causes. For example, ``outliers cause unexpected outcomes'' is just one of the
many possible hypotheses, which offers very limited expressivity. Second, we
need to \emph{verify} the hypotheses to confirm or refute them, which enables
us to pinpoint the actual root causes, eliminating false positives.

\subsubsection*{Data profile as root cause.} Towards solving the first
challenge, our observation is that data-driven systems often function properly
for certain datasets, but malfunction for others. Such malfunction is often
rooted in certain \emph{properties} of the data, which we call \emph{data
profiles}, that distinguish passing and failing datasets. Examples include size
of a dataset, domains and ranges of attribute values, correlations between
attribute pairs, conditional independence~\cite{DBLP:conf/sigmod/YanSZWC20},
functional dependencies and their variants~\cite{papenbrock2015functional,
koudas2009metric, DBLP:conf/icde/FanGLX09, ilyas2004cords,
caruccio2016discovery}, and other more complex data
profiles~\cite{song2011differential, DBLP:journals/vldb/LangerN16,
papenbrock2015divide, DBLP:journals/pvldb/ChuIP13}.

\subsubsection*{Oracle-guided root cause identification.} Our second
observation is that if we have access to an \emph{oracle} that can indicate
whether the system functions desirably or not, we can verify our hypotheses.
Access to an oracle allows us to precisely isolate the correct root causes of
the undesirable malfunction from a set of candidate causes. Here, an oracle is
a mechanism that can characterize whether the system functions properly over
the input data. The definition of proper functioning is application-specific;
for example, achieving a certain accuracy may indicate proper functioning for
an ML pipeline. Such oracles are often available in many practical settings,
and have been considered in prior work~\cite{bugdoc, aid}.

\smallskip

\paragraph{Solution sketch.} In this paper, we propose \system, a framework
that identifies and exposes data profiles that cause a data-driven system to
malfunction. Our framework involves two main components: (1)~an
intervention-based mechanism that alters the profiles of a dataset, and (2)~a
mechanism that speeds up analysis by carefully selecting appropriate
interventions. Given a scenario where a system malfunctions (fails) over a
dataset but functions properly (passes) over another, \system focuses on the
\emph{discriminative} profiles, i.e., data profiles that significantly differ
between the two datasets. \system's intervention mechanism modifies the
``failing'' dataset to alter one of the discriminative profiles; it then
observes whether this intervention causes the system to perform desirably, or
the malfunction persists. \system speeds up this analysis by favoring
interventions on profiles that are deemed more likely causes of the
malfunction. To estimate this likelihood, we leverage three properties of a
profile:
(1)~\emph{coverage}: the more tuples an intervention affects, the more likely
it is to fix the system behavior,
(2)~\emph{discriminating power}: the bigger the difference between the failing
and the passing datasets over a profile, the more likely that the profile is a
cause of the malfunction, and
(3)~\emph{attribute association}: if a profile involves an attribute that is
also involved with a large number of other discriminative profiles, then that
profile has high likelihood to be a root cause. This is because altering such a
profile is likely to passively repair other discriminative profiles as a
side-effect (through the associated attribute).
We also provide a group-testing-based technique that allows \emph{group
intervention}, which helps expedite the root-cause analysis further.

\smallskip

\paragraph{Scope.} In this work, we assume knowledge of the classes of
(domain-specific) data profiles that encompass the potential root causes. E.g.,
in Example~\ref{ex:one}, we assume the knowledge that correlation between
attribute pairs and disparity between the conditional probability distributions
(the probability of belonging to a certain gender, given price of items bought)
are potential causes of malfunction. This assumption is realistic because:
(1)~For a number of tasks there exists a well-known set of relevant profiles:
e.g., class imbalance and correlation between sensitive and non-sensitive
attributes are common causes of unfairness in
classification~\cite{bellamy2018ai}; and violation of conformance
constraints~\cite{fariha2021conformance}, missing values, and
out-of-distribution tuples are well-known causes of ML model's performance
degradation. (2)~Domain experts are typically aware of the likely class of data
profiles for the specific task at hand and can easily provide this additional
knowledge as a \emph{conservative} approximation, i.e., they can include extra
profiles just to err on the side of caution. Notably, this assumption is also
extremely common in software debugging techniques~\cite{aid,
liblit2005scalable, Zheng2006}, which rely on the assumption that the
``predicates'' (traps to extract certain runtime conditions) are expressive
enough to encode the root causes, and software
testing~\cite{mesbah2011invariant}, validation~\cite{kumar2013verification},
and verification~\cite{henzinger2003software} approaches, which rely on the
assumption that the test cases, specifications, and invariants reasonably cover
the codebase and correctness constraints.

\looseness-1 To support a data profile, \system further needs the corresponding
mechanisms for discovery and intervention. In this work, we assume knowledge of
the profile discovery and intervention techniques, as they are orthogonal to
our work. Nevertheless, we discuss some common classes of data profiles
supported in \system and the corresponding discovery and intervention
techniques. For data profile discovery, we rely on prior work on pattern
discovery~\cite{pattern}, statistical-constraint
discovery~\cite{DBLP:conf/sigmod/YanSZWC20}, data-distribution
learning~\cite{hellerstein2008quantitative}, knowledge-graph-based concept
identification~\cite{galhotra2019automated}, etc. While our evaluation covers
specific classes of data profiles (for which there exist efficient discovery
techniques), our approach is generic and works for any class of data profiles,
as long as the corresponding discovery and intervention techniques are
available.

\smallskip

\paragraph{Limitations of prior work.} \looseness-1 To find potential issues in
data, Dagger~\cite{DBLP:conf/cidr/RezigCSSMTOS20,
DBLP:journals/pvldb/RezigBTO0MMS20} provides data debugging primitives for
human-in-the-loop interactions with data-driven computational pipelines. Other
explanation-centric efforts~\cite{DBLP:conf/sigmod/WangDM15,
Bailis:2017:MPA:3035918.3035928, Chirigati:2016:DPM:2882903.2915245,
GebalyFGKS14} report salient properties of historical data based only on
observations. In contrast with observational techniques, the presence of an
oracle allows for interventional techniques~\cite{bugdoc} that can query the
oracle with additional, system-generated test cases to identify root causes of
system malfunction more accurately. One such approach is
CheckCell~\cite{DBLP:conf/oopsla/BarowyGB14}, which presents a ranked list of
cells of data rows that unusually affect output of a given target function.
CheckCell uses a fine-grained approach: it removes one cell of the data at a
time, and observes changes in the output distribution. While it is suitable for
small datasets, where it is reasonable to expect a human-in-the-loop paradigm
to fix cells one by one, it is not suitable for large datasets, where no
individual cell is significantly responsible, rather, a holistic property of
the entire dataset (profile) causes the problem.

Interpretable machine learning is related to our problem, where the goal is to
explain behavior of machine-learned models. However, prior work on
interpretable machine learning~\cite{lime, anchors} typically provide
\emph{local} (tuple-level) explanations, as opposed to \emph{global}
(dataset-level) explanations. While some approaches provide feature importance
as a global explanation for model behavior~\cite{casalicchio2018visualizing},
they do not model feature interactions as possible explanations.

\looseness-1 Software testing and debugging techniques~\cite{Gulzar2018,
Attariyan2011, Attariyan2012, chen2002pinpoint, fraser@tse2013,
godefroid@ndss2008, holler@uss2012, DBLP:conf/icse/JohnsonB20, Zheng2006,
liblit2005scalable} are either application-specific, require user-defined test
suites, or rely only on observational data. The key contrast between software
debugging and our setting is that the former focuses on white-box programs:
interventions, runtime conditions, program invariants, control-flow graphs,
etc., all revolve around program source code and execution traces. Unlike
programs, where lines have logical and semantic connections, tuples in data do
not have similar associations. Data profiles significantly differ in their
semantics, and discovery and intervention techniques from program profiles,
and, thus, techniques for program profiling do not trivially apply here. We
treat data as a first-class citizen in computational pipelines, while
considering the program as a black box.

\smallskip

\paragraph{Contributions.} In this paper, we make the following contributions:

\begin{itemize}
	
	\item We formalize the novel problem of identifying root causes (and fixes)
of the disconnect between data and data-driven systems in terms of data
profiles (and interventions). (Sec~\ref{sec:problem})

	\item We design a set of data profiles that are common root causes of
data-driven system malfunctions, and discuss their discovery and intervention
techniques based on available technology. (Sec~\ref{sec:profile})

	\item We design and develop a novel interventional approach to pinpoint
causally verified root causes. The approach leverages a few properties of the
data profiles to efficiently explore the space of candidate root causes with a
small number of interventions. Additionally, we develop an efficient
group-testing-based algorithm that further reduces the number of required
interventions. (Sec~\ref{sec:algorithm})

	\item We evaluate \system on three real-world applications, where
data profiles are responsible for causing system malfunction, and demonstrate
that \system successfully explains the root causes with a very small number
of interventions (< $5$). Furthermore, \system requires $10$--$1000\times$ fewer
interventions when compared against two state-of-the-art techniques for
root-cause analysis: BugDoc~\cite{bugdoc} and Anchors~\cite{anchors}. Through
an experiment over synthetic pipelines, we further show that the number of
required interventions by \system increases sub-linearly with the number of
discriminative profiles, thanks to our group-testing-based approach.
(Sec~\ref{sec:experiment})
	
\end{itemize}

\section{Preliminaries \& Problem Definition}\label{sec:problem}
%

\def\pvt{PVT\xspace}

In this section, we first formalize the notions of system malfunction and data
profile, its violation, and transformation function used for intervention. We
then proceed to define explanation (cause and corresponding fix) of system
malfunction and formulate the problem of data-profile-centric explanation of
system malfunction.

\smallskip

\paragraph{Basic notations.} We use $\mathcal{R} (A_1, A_2, \dots, A_m)$ to
denote a relation schema over $m$ attributes, where $A_i$ denotes the $i^{th}$
attribute. We use $\Dom_i$ to denote the domain of attribute $A_i$. Then the
set ${\DDom}^m=\Dom_1\times \cdots\times \Dom_m$ specifies the domain of all
possible tuples. A dataset $D\subseteq {\DDom}^m$ is a specific instance of the
schema $\mathcal{R}$. We use $t \in\DDom^m$ to denote a tuple in the schema
$\mathcal{R}$. We use $t.A_i \in \Dom_i$ to denote the value of the attribute
$A_i$ of the tuple $t$ and use $D.A_j$ to denote the multiset of values all
tuples in $D$ take for attribute $A_j$.

\subsection{Quantifying System Malfunction}
To measure how much the system malfunctions over a dataset, we use the
\emph{malfunction score}.

\begin{definition}[Malfunction score]
	 Let $D \subseteq \DDom^m$ be a dataset, and $S$ be a system operating on
	 $D$. The malfunction score $m_S(D) \in [0, 1]$ is a real value that
	 quantifies how much $S$ \emph{malfunctions} when operating on $D$.
\end{definition}

The malfunction score $m_S(D)=0$ indicates that $S$ functions properly over $D$
and a higher value indicates a higher degree of malfunction, with $1$
indicating extreme malfunction.
A threshold parameter $\tau$ defines the acceptable degree of malfunction and
translates the continuous notion of malfunction to a Boolean value. If $m_S(D)
\le \tau$, then $D$ is considered to pass with respect to $S$; otherwise, there
exists a mismatch between $D$ and $S$, whose cause (and fix) we aim to expose.

\begin{example}
	For a binary classifier, its misclassification rate (additive inverse of
accuracy) over a dataset can be used as a malfunction score. Given a dataset
$D$, if a classifier $S$ makes correct predictions for tuples in $D' \subseteq
D$ , and incorrect predictions for the remaining tuples, then $S$ achieves
accuracy $\frac{\lvert D' \rvert}{\lvert D \rvert}$, and, thus, $m_S(D) =
1-\frac{\lvert D' \rvert}{\lvert D \rvert}$.
\end{example}

\begin{example} 
	In fair classification, we can use disparate impact~\cite{ibmaif360}, which
is defined by the ratio between the number of tuples with favorable outcomes
within the unprivileged and the privileged groups, to measure malfunction.
\end{example}

\subsection{Profile-Violation-Transformation (\pvt)} 
Once we detect existence of a mismatch, the next step is to investigate its
cause. We characterize the issues in a dataset that are responsible for the
mismatch between the dataset and the system using \emph{data profiles}.
Structure or schema of data profiles is given by profile \emph{templates},
which contains holes for parameters. Parameterizing a profile template gives us
a \emph{concretization} of the corresponding profile ($P$). Given a dataset
$D$, we use existing data-profiling techniques to find out parameter values to
obtain concretized data profiles, such that $D$ satisfies the concretized
profiles. To evaluate how much a dataset $D$ satisfies or violates a data
profile, we need a corresponding violation function ($V$). Violation functions
provide \emph{semantics} of the data profiles. Finally, to alter a dataset $D$,
with respect to a data profile and the corresponding violation function, we
need a transformation function ($T$). Transformation functions provide the
intervention mechanism to alter data profiles of a dataset and suggest fix to
remove the cause of malfunction. \system requires the following three
components over the schema $\langle$\textbf{P}rofile, \textbf{V}iolation
function, \textbf{T}ransformation function$\rangle$, \pvt in short:

\begin{enumerate}
	
	\item $P$: a (concretized) profile along with its parameters, which follows
the schema $\langle$profile type, parameters$\rangle$.
	
	\item$V(D, P)$: a violation function that computes how much the dataset $D$
violates the profile $P$ and returns a violation score.

	\item $\T (D, P, V)$: a transformation function that transforms the dataset
$D$ to another dataset $D'$ such that $D'$ no longer violates the profile $P$
with respect to the violation function $V$. (When clear from the context, we
omit the parameters $P$ and $V$ when using the notation for transformation
functions.)

\end{enumerate} 

\smallskip

For a \pvt triplet $X$, we define $X_P$ as its profile, $X_V$ as the violation
function and $X_\T$ as the transformation function. We provide examples and
additional discussions on data profiles, violation functions, and
transformation functions in Section~\ref{sec:profile}.


\subsubsection{Data Profile}

Intuitively, data profiles encode dataset characteristics. They can refer to a
single attribute (e.g., mean of an attribute) or multiple attributes (e.g.,
correlation between a pair of attributes, functional dependencies, etc.).

\begin{definition}[Data Profile]
		
	Given a dataset $D$, a data profile $P$ denotes properties or constraints
that tuples in $D$ (collectively) satisfy.
	
\end{definition}

\subsubsection{Profile Violation Function} 
To quantify the degree of violation a dataset incurs with respect to a data 
profile, we use a \emph{profile violation function} that returns a numerical 
violation score.

\begin{definition}[Profile violation function]
	Given a dataset $D$ and a data profile $P$, a \emph{profile violation
function} $V(D, P) \mapsto [0, 1]$ returns a real value that quantifies how much
$D$ violates $P$.
\end{definition}

$V(D, P) = 0$ implies that $D$ fully complies with $P$ (does not violate it at 
all). In contrast, $V(D, P) > 0$ implies that $D$ violates $P$. The higher the 
value of $V(D, P)$, the higher the profile violation.

\begin{figure*}
	\centering
	\renewcommand{\arraystretch}{1.2}
	\resizebox{\textwidth}{!}{
	\begin{tabular}{
		l
		>{\columncolor{vlightgray}}l
		>{\columncolor{vlightgray}}l
		>{\columncolor{vlightgray}}p{1.5cm}
		>{\columncolor{vlightgray}}p{3.3cm}
		>{\columncolor{vlightgray}}p{3.3cm}
		>{\columncolor{vlightgray}}p{4.3cm}
		>{\columncolor{vlightgray}}l
	}

		\toprule
		& \multicolumn{1}{l}{}
		& \multicolumn{1}{l}{\textbf{Profile}}
		& \multicolumn{1}{l}{\textbf{Data type}}
		& \multicolumn{1}{l}{\textbf{Discovery over $D$}}
		& \multicolumn{1}{l}{\textbf{Interpretation}}
		& \multicolumn{1}{l}{\textbf{Violation by $D$}} 								
		& \multicolumn{1}{l}{\textbf{Transformation function}} 	  			
		\\
		 
 		\midrule
		
		\rowcolor{white}
		& $\mathbf{1}$
		& \begin{tabular}{@{}l} $\langle$\pl{$\domain$}, $A_j$, $\mathbb{S} \rangle$\end{tabular}
		& \begin{tabular}{@{}l}Categorical\end{tabular}			
		& 
				\begin{tabular}{@{}l}
					$\mathbb{S} = \displaystyle\bigcup_{t \in D} \{ t.A_j \}$
				\end{tabular}
		& \begin{tabular}{@{}p{3.3cm}}Values are drawn from a specific domain.\end{tabular}
		& \begin{tabular}{@{}l}$\frac{\sum_{t\in D} \llbracket t.A_j \;  \not\in \; \mathbb{S}\rrbracket}{\lvert D \rvert}$\end{tabular}								
		& \begin{tabular}{@{}p{4.3cm}}Map  values outside $\mathbb{S}$ to values in $\mathbb{S}$ using domain knowledge.\end{tabular}
		\\

		& $\mathbf{2}$
		& \begin{tabular}{@{}l}$\langle$\pl{$\domain$}, $A_j$, $\mathbb{S}\rangle$\end{tabular}	
		& \begin{tabular}{@{}l}Numerical\end{tabular}
		& 
			\begin{tabular}{@{}l@{$\;\;$}ll}
			$\mathbb{S}$ & = & $[\mathtt{lb}, \mathtt{ub}]$,	where\\
			$\mathtt{lb}$ & = & $\displaystyle\min_{t \in D} t.A_j$\\
			$\mathtt{ub}$ & = & $\displaystyle\max_{t \in D} t.A_j$
		  \end{tabular}
		& \begin{tabular}{@{}p{3.3cm}}Values lie within a bound.\end{tabular}
		& \begin{tabular}{@{}l}$\frac{\sum_{t\in D} \llbracket t.A_j \;  \not\in \; \mathbb{S}\rrbracket}{\lvert D \rvert}$\end{tabular}
		& \begin{tabular}{@{}p{4.3cm}}
			(1)~Use monotonic linear transformation and transform all values.\\
			(2)~Use winsorization techniques to replace the violating values only.
		\end{tabular}														 	 										
		\\

		\rowcolor{white}
		\multirow{-7}{*}{\turnrow{\makecell[l]{\textbf{strict}}}}
		& $\mathbf{3}$
		& \begin{tabular}{@{}l}$\langle$\pl{$\domain$}, $A_j$, $\mathbb{S} \rangle$\end{tabular}
		& \begin{tabular}{@{}l}Text\end{tabular}
		& 
		\begin{tabular}{@{}l}
			$\mathbb{S} = [t \in \Dom_j \mid t \models \mathbb{P}]$,\\
			where $\mathbb{P}$ is a regex\\
			over $D.A_j$ learned via\\
			pattern discovery~\cite{pattern}
		\end{tabular}
		& \begin{tabular}{@{}p{3.3cm}}Values satisfy a regular expression or length of values lie within a bound.\end{tabular}										 	
		& \begin{tabular}{@{}l}$\frac{\sum_{t\in D} \llbracket t.A_j \;  \not\in \; \mathbb{S}\rrbracket}{\lvert D \rvert}$\end{tabular}
		& \begin{tabular}{@{}p{4.3cm}}Minimally alter data to satisfy regular expression. For example, insert (remove) characters to increase (reduce) text length.\end{tabular}
		\\

		\toprule
		\toprule
		
		& $\mathbf{4}$
		& \begin{tabular}{@{}l}$\langle$\pl{$\outlier$}, $A_j$, {$O$}, $\theta \rangle$\end{tabular}
		& \begin{tabular}{@{}l}All\end{tabular}
		& 
				\begin{tabular}{@{}l}
					$\theta = \frac{\sum_{t\in D} \llbracket O(D.A_j, t.A_j) \rrbracket}{\lvert D \rvert}$,\\
					where $O$ is learned from\\
					$D.A_j$'s distribution~\cite{hellerstein2008quantitative}
				\end{tabular}
		& \begin{tabular}{@{}p{3.3cm}}Fraction of outliers within an attribute does not exceed a threshold.\end{tabular}
		& \begin{tabular}{@{}l}$\max\Big(0, \frac{\sum_{t\in D} \llbracket O(D.A_j, t.A_j) \rrbracket - \theta \cdot \lvert D \rvert }{\lvert D \rvert \cdot (1 - \theta)} \Big)$\end{tabular}
		& \begin{tabular}{@{}p{4.3cm}}
		(1)~Replace outliers with the expected value (mean, median, mode) of the attribute.\\
		(2)~Map all values above (below) the maximum (minimum) limit with highest (lowest) valid value.
		\end{tabular}										
		\\

		\rowcolor{white}
		& $\mathbf{5}$
		& \begin{tabular}{@{}l}$\langle$\pl{$\missing$}, $A_j$, $\theta \rangle$\end{tabular}
		& \begin{tabular}{@{}l}All\end{tabular}
		& \begin{tabular}{@{}l}$\theta = \frac{\sum_{t\in D} \llbracket t.A_j = \mathtt{NULL} \rrbracket}{\lvert D \rvert}$\end{tabular}
		& \begin{tabular}{@{}p{3.3cm}}Fraction of missing values within an attribute does not exceed a threshold.\end{tabular}								 	
		& \begin{tabular}{@{}l}$\max\Big(0, \frac{\sum_{t\in D} \llbracket t.A_j = \mathtt{NULL}  \rrbracket - \theta \cdot \lvert D \rvert }{\lvert D \rvert \cdot (1 - \theta)} \Big)$\end{tabular}
		& \begin{tabular}{@{}p{4.3cm}}Use missing value imputation techniques.\end{tabular}
		\\
		
		\multirow{-8}{*}{\turnrow{\makecell[l]{\textbf{thresholded by data coverage}}}}
		& $\mathbf{6}$
		& \begin{tabular}{@{}l}$\langle$\pl{$\selectivity$}, $\mathbb{{P}}, \theta \rangle$\end{tabular}
		& \begin{tabular}{@{}l}All\end{tabular}
		& \begin{tabular}{@{}l}$\theta = \frac{\lvert \sigma_\mathbb{P}(D) \rvert}{\lvert D \rvert}$\end{tabular}
		& \begin{tabular}{@{}p{3.3cm}}Fraction of tuples satisfying a given constraint (selection predicate) does not exceed a threshold.\end{tabular}
		& \begin{tabular}{@{}l}$\max\Big(0, \frac{\lvert \sigma_\mathbb{P}(D) \rvert - \theta \cdot \lvert D \rvert }{\lvert D \rvert \cdot (1 - \theta)} \Big)$\end{tabular}
		& \begin{tabular}{@{}p{4.3cm}}Undersample tuples that satisfy the predicate $\mathbb{P}$.\end{tabular}
		\\
		
		\toprule
		\toprule
	
		\rowcolor{white}
		& $\mathbf{7}$
		& \begin{tabular}{@{}l}$\langle$\pl{$\correlation$}, $A_j$, $A_k$, $\alpha \rangle$\end{tabular}
		& \begin{tabular}{@{}l}Categorical\end{tabular}
		& 
			\begin{tabular}{@{}l}
				$\alpha$ denotes Chi-squared\\
				statistic between $D.A_j$\\
				and $D.A_k$
			\end{tabular}
		& \begin{tabular}{@{}p{3.3cm}} $\chi^2$ statistic between a pair of attributes is below a threshold	with a p-value $\le$ 0.05.\end{tabular}
		& 
			\begin{tabular}{@{}l}
				$1 - e^{-\max(0, \chi^2(D.A_j, D.A_k) - \alpha)}$\\
			\end{tabular}
		& \begin{tabular}{@{}p{4.3cm}}Modify attribute values to remove/reduce dependence.\end{tabular}
		\\

		& $\mathbf{8}$		
		& \begin{tabular}{@{}l}$\langle$\pl{$\correlation$}, $A_j$, $A_k$, $\alpha \rangle$\end{tabular}
		& \begin{tabular}{@{}l}Numerical\end{tabular}
		& 
			\begin{tabular}{@{}l}
				$\alpha$ denotes Pearson\\
				correlation coefficient\\
				between $D.A_j$ and $D.A_k$
			\end{tabular}
		& \begin{tabular}{@{}p{3.3cm}} PCC between a pair attributes is below a threshold with a p-value $\le$ 0.05.\end{tabular}
		& 
			\begin{tabular}{@{}l}
				$\max\big(0, \frac{\lvert \mathtt{PCC}(D.A_j, D.A_k) \rvert - \lvert \alpha \rvert}{1 - \lvert \alpha \rvert}\big)$ 	
			\end{tabular}
		& \begin{tabular}{@{}p{4.3cm}}Add noise to remove/reduce\\dependence between attributes.\end{tabular}
		\\

		\rowcolor{white}
		\multirow{-8}{*}{\turnrow{\makecell[l]{\textbf{thresholded by parameter}}}}
		& $\mathbf{9}$
		& \begin{tabular}{@{}l}$\langle$\pl{$\correlation$}, $A_j$, $A_k, \alpha \rangle$\end{tabular}
		& \begin{tabular}{@{}l}Categorical,\\ numerical\end{tabular}
		& \begin{tabular}{@{}l}Learn causal graph and  \\causal coefficients ($\alpha$)\\using TETRAD~\cite{scheines1998tetrad}\end{tabular}
		& \begin{tabular}{@{}p{3.3cm}}A causal relationship between a pair of attributes is unlikely (with causal coefficient less than $\alpha$).\end{tabular}
		& \begin{tabular}{@{}l}$\max\big(0, \frac{|\mathtt{coeff}( A_j, A_k)| - \alpha}{1 - \alpha}\big)$\end{tabular}	
		& \begin{tabular}{@{}p{4.3cm}}Change data distribution to modify the causal relationship.\end{tabular}									 											
		\\

			%
		%
		\bottomrule
	
	\end{tabular}
	}
	\vspace{-3mm}
\caption{A list of \pvt triplets that we consider in this paper, their syntax, and semantics.}
	\vspace{-2mm}
\label{fig:pvt}
\end{figure*}

\subsubsection{Transformation Function} In our work, we assume knowledge of a
passing dataset for which the system functions properly, and a failing dataset
for which the system malfunctions. Our goal is to identify which profiles of
the failing dataset caused the malfunction. We seek answer to the question: how
to ``fix'' the issues within the failing dataset such that the system no longer
malfunctions on it (mismatch is resolved)? To this end, we apply
\emph{interventional causal reasoning}: we intervene on the failing
dataset by altering its attributes such that the profile of the altered dataset
matches the corresponding correct profile of the passing dataset. 
To perform intervention, we need \emph{transformation functions} with the 
property that it should push the failing dataset ``closer'' to the passing 
dataset in terms of the profile that we are  interested to alter. More 
formally, after the transformation, the profile violation score should 
decrease.

\begin{definition}[Transformation function]
Given a dataset $D$, a data profile $P$, and a violation function $V$, a
transformation function $\T (D, P, V) \mapsto 2^{\DDom^m}$ alters $D$ to
produce $D'$ such that $V(D', P) = 0$.
\end{definition}
A dataset can be transformed by applying a series of transformation functions,
for which we use the composition operator ($\circ$).

\begin{definition}[Composition of transformations]
	\begin{sloppypar} Given a dataset $D$, and two \pvt triplets $X$ and $Y$,
$(X_\T \circ Y_\T)(D) = X_\T(Y_\T(D))$. Further, if $D'' = (X_\T \circ
Y_\T)(D)$, then $ X_V(D'', X_P) = Y_V(D'', Y_P) = 0$. \end{sloppypar}
\end{definition}

\subsection{Problem Definition} We expose a set of \pvt triplets for explaining
the system malfunction. The explanation contains both the \emph{cause} and the
corresponding \emph{fix}: profile within a \pvt triplet indicates the cause of
system malfunction with respect to the corresponding transformation function,
which suggests the fix.

\begin{definition}[Explanation of system malfunction]
	Given 
	\begin{enumerate}
		\item a system $S$ with a mechanism to compute $m_S(D)$ $\forall D \subseteq \DDom^m$, 
		\item an allowable malfunction threshold $\tau$,
		\item a passing dataset $D_{\mathit{pass}}$ for which $m_S(D_{\mathit{pass}}) \le \tau$, 
		\item a failing dataset $D_{\mathit{fail}}$ for which $m_S(D_{\mathit{fail}}) > \tau$, and
		
		\item \begin{sloppypar} a set of candidate \pvt triplets $\P$ such that $\forall X \in \P \;\;$  $X_V(D_{\mathit{pass}}, X_P) = 0 \wedge X_V(D_{\mathit{fail}}, X_P) > 0$,\end{sloppypar}
	\end{enumerate}
	the explanation of the malfunction of $S$ for $D_{\mathit{fail}}$, but not
for $D_{\mathit{pass}}$, is a set of \pvt triplets $\E \subseteq {\P}$
such that $m_S((\circ_{X \in \E}X_\T)(D_{\mathit{fail}})) \le \tau$.
\end{definition}

\looseness-1 Informally, $\E$ explains the cause: why $S$ malfunctions for
$D_{\mathit{fail}}$, but not for $D_{\mathit{pass}}$. More specifically,
\emph{failing to satisfy} the profiles of the \pvt triplets in $\E$ are the
causes of malfunction.
Furthermore, the transformation functions of the \pvt triplets in $\E$ suggest
the fix: how can we repair $D_{\mathit{fail}}$ to eliminate system malfunction.
However, there could be many possible such $\E$ and we seek a \emph{minimal}
set $\E$ such that transformation for every $X \in \E$ is necessary to bring
down the malfunction score below the threshold $\tau$.

\begin{definition}[Minimal explanation of system malfunction]
	Given a system $S$ that malfunctions for $D_{\mathit{fail}}$ and an allowable
malfunction threshold $\tau$, an explanation $\E$ of $S$'s malfunction for $D_{\mathit{fail}}$ is
minimal if $ \; \forall\P' \subset \E \; m_S((\circ_{X \in \P'}X_\T)(D_{\mathit{fail}})) >
\tau$.
\end{definition}

\looseness-1 Note that there could be multiple such minimal explanations and we
seek any one of them, as any minimal explanation exposes the causes of mismatch
and suggests minimal fixes.

\begin{problem}[Discovering explanation of mismatch between data and system]
	 Given a system $S$ that malfunctions for $D_{\mathit{fail}}$ but functions
	 properly for $D_{\mathit{pass}}$, the problem of discovering the explanation
	 of mismatch between $D_{\mathit{fail}}$ and $S$ is to find a minimal
	 explanation that captures (1)~the cause why $S$ malfunctions for
	 $D_{\mathit{fail}}$ but not for $D_{\mathit{pass}}$ and (2)~how to repair
	 $D_{\mathit{fail}}$ to remove the malfunction.
\end{problem}

\section{Data Profiles, Violation Functions, \& Transformation Functions}\label{sec:profile}
%

\looseness-1 We now provide an overview of the data profiles we consider, how
we discover them, how we compute the violation scores for a dataset w.r.t. a
data profile, and how we apply transformation functions to alter profiles of a
dataset. While a multitude of data-profiling primitives exist in the
literature, we consider a carefully chosen subset of them that are particularly
suitable for modeling issues in data that commonly cause malfunction or failure
of a system. We focus on profiles that, by design, can better ``discriminate''
a pair of datasets as opposed to ``generative'' profiles (e.g., data
distribution) that can profile the data better, but nonetheless are less useful
for the task of discriminating between two datasets. However, the \system
framework is generic, and other profiles can be plugged into it.

\looseness-1 As discussed in Section~\ref{sec:problem}, a \pvt triplet
encapsulates a profile, and corresponding violation and transformation
functions. Figure~\ref{fig:pvt} provides a list of profiles along with the data
types they support, how to learn their parameters from a given dataset, how to
interpret them intuitively, and the corresponding violation and transformation
functions. In this work, we assume that a profile can be associated with
multiple transformation functions (e.g., rows 2 and 4), but each transformation
function can be associated with at most one profile. This assumption helps us
to blame a unique profile as cause of the system malfunction when at least one
of the transformation functions associated with that profile is verified to be
a fix.

\pvt triplets can be classified in different ways. Based on the strictness of
the violation function, they can be classified as follows:
\begin{itemize} 
	
	\item \emph{Strict}: \emph{All} tuples are expected to satisfy the profile
(rows 1--3).

	\item \emph{Thresholded by data coverage:} Certain fraction ($\theta$) of
data tuples are allowed to violate the profile (rows 4--6).

	\item \emph{Thresholded by a parameter:} Some degree of violation is
allowed with respect to a specific parameter ($\alpha$) (rows 7--9).

\end{itemize}

\smallskip

Further, \pvt triplets can be classified in two categories based on the nature
of the transformation functions: 

\begin{itemize}

	\item \emph{Local} transformation functions can transform a tuple in
isolation without the knowledge of how other tuples are being transformed
(e.g., rows 1--3). Some local transformation functions only transform the
violating tuples (e.g., row 2, transformation (2)), while others transform all
values (e.g., row 2, transformation (1)). For instance, in case of unit
mismatch (kilograms vs. lbs), it is desirable to transform all values and not
just the violating ones.

	\item \emph{Global} transformation functions are holistic, as they need the
knowledge of how other tuples are being transformed while transforming a tuple
(e.g., rows 6 and 9).

\end{itemize}

\begin{example}
	 \textbf{\domain} requires two parameters: (1)~an attribute $A_j \in
	 \mathcal{R}(D)$, and (2)~a set $\mathbb{S}$ specifying its domain. A dataset
	 $D$ satisfies \pl{$\langle \domain, A_j, \mathbb{S} \rangle$} if $\;\forall t
	 \in D \;\; t.A_j \in \mathbb{S}$. The profile \pl{$\langle \domain, A_j,
	 \mathbb{S} \rangle$} is minimal w.r.t. $D$ if $\;\not\exists \mathbb{S}'
	 \subset \mathbb{S}$ s.t. $D$ satisfies the profile \pl{$\langle \domain, A_j,
	 \mathbb{S}' \rangle$}. The technique for discovering a domain $\mathbb{S}$
	 varies depending on the data type of the attribute. Rows 1--3 of
	 Figure~\ref{fig:pvt} show three different domain-discovery techniques for
	 different data types.

	 \pplfail (Figure~\ref{fig:faildb}) satisfies \pl{$\langle \domain,
	 \mathtt{gender}, \{F, M\} \rangle$}, as all tuples draw values from $\{F,
	 M\}$ for the attribute \texttt{gender}. Our case studies of \texttt{Sentiment
	 Prediction} and \texttt{Cardiovascular Disease Prediction} show the
	 application of the profile {\domain} (Section~\ref{sec:experiment}).
	
\end{example}

\begin{example}
	
	 \textbf{\outlier} requires three parameters: (1)~an attribute $A_j \in
	 \mathcal{R}(D)$, (2)~an outlier detection function $O (A, a) \mapsto \{
	 \mathtt{True}, \mathtt{False}\}$ that returns $\mathtt{True}$ if $a$ is an
	 outlier w.r.t. the values within $A$, and $\mathtt{False}$ otherwise, and
	 (3)~a threshold $\theta \in [0, 1]$. A dataset $D$ satisfies \pl{$\langle
	 \outlier, A_j, O, \theta \rangle$} if the fraction of outliers within the
	 attribute $A_j$---according to $O$---does not exceed $\theta$. Otherwise, we
	 compute how much the fraction of outliers exceeds the allowable fraction of
	 outliers ($\theta$) and then normalize it by dividing by $(1 - \theta)$. The
	 profile \pl{$\langle \outlier, A_j, O, \theta \rangle$} is minimal if
	 $\;\not\exists\theta' < \theta$ s.t. $D$ satisfies \pl{$\langle \outlier,
	 A_j, O, \theta' \rangle$}.

	 An outlier detection function $O_{1.5}$ identifies values that are more
	 than $1.5$ standard deviation away from the mean as outliers. In \pplfail,
	 \texttt{age} has a mean $34.5$ and a standard deviation $11.78$. According to
	 $O_{1.5}$, only $t_3$---which is $0.1$ fraction of the tuples---is an outlier
	 in terms of \texttt{age} as $t_3$'s age $60 > (34.5 + 1.5 \times 11.78) =
	 52.17$. Therefore, \pplfail satisfies \pl{$\langle \outlier, \mathtt{age},
	 O_{1.5}, 0.1 \rangle$}.
	
\end{example}

\begin{example}
	\noindent \textbf{\correlation} requires three parameters: two attributes
$A_j, A_k \in \mathcal{R}(D)$, and a real value $\alpha$. A dataset $D$
satisfies the profile \pl{$\langle \correlation, A_j, A_k, \alpha \rangle$} if
the dependency between $D.A_j$ and $D.A_k$ does not exceed $\alpha$. Different
techniques exist to quantify the dependency and rows 6--9 of
Figure~\ref{fig:pvt} show three different ways to model dependency, where the
first two are correlational and the last one is causal.

	\pl{$\langle \correlation, \mathtt{race}, \mathtt{high\_expenditure},0.67
\rangle$} is satisfied by \pplfail using the \pvt triplet of row 7, as
$\chi^2$-statistic between \texttt{race} and \texttt{high\_expenditure} over
\pplfail is $0.67$. We show the application of the profile {\correlation} in
our case study involving the task of \texttt{Income Prediction} in
Section~\ref{sec:experiment}.

\end{example}

While the profiles in Figure~\ref{fig:pvt} are defined over the entire data,
analogous to conditional functional dependency~\cite{DBLP:conf/icde/FanGLX09},
an extension to consider is \emph{conditional} profiles, where only a subset of
the data is required to satisfy the profiles.

\begin{figure}
	\centering
	\resizebox{1\columnwidth}{!}{
	\setlength{\tabcolsep}{1mm}
	\begin{tabular}{llclcccc}
		\toprule
		\textbf{id} & \textbf{name}		& \textbf{gender}  & \textbf{age}	 & \textbf{race} 	& \textbf{zip code} 	& \textbf{phone} & \makecell[c]{\textbf{high} \textbf{expenditure}}			\\
		\midrule
		$t_1$ 		&	Shanice Johnson		& F   		& 45	& A  			& 01004    			& 2088556597 	& no  \\
		$t_2$ 		&	DeShawn Bad			& M 			& 40	& A    			& 01004    			& 2085374523 	& no  \\
		$t_3$ 		&	Malik Ayer 			& M   		& 60	& A    			& 01005   			& 2766465009	& no  \\
		$t_4$ 		&	Dustin Jenner		& M   		& 22	& W    			& 01009    			& 7874891021	& yes \\
		$t_5$ 		&	Julietta Brown		& F   		& 41	& W    			& 01009    			& 				& yes \\
		$t_6$ 		&	Molly Beasley		& F   		& 32	& W   			&    				& 7872899033	& no  \\
		$t_7$ 		&	Jake Bloom			& M   		& 25	& W    			& 01101    			& 4047747803	& yes \\
		$t_8$ 		&	Luke Stonewald		& M   		& 35	& W   			& 01101    			& 4042127741	& yes \\
		$t_9$ 		&	Scott Nossenson 	& M  		& 25	& W  			& 01101    			& 				& yes \\
		$t_{10}$ 	&	Gabe Erwin			& M   		& 20	& W 			&    				& 4048421581 	& yes \\
		\bottomrule
	\end{tabular}
	}
	\vspace{-2mm}
	\caption{A sample dataset \pplfail with $10$ entities. A logistic
regression classifier trained over this dataset discriminates against African Americans (\texttt{race} = `A') and women (\texttt{gender} = `F') (Example~\ref{ex:one}).}
\vspace{-2mm}
	\label{fig:faildb}
\end{figure}

\begin{figure}[t]
	\centering
	\resizebox{1\columnwidth}{!}{
	\setlength{\tabcolsep}{1mm}
	\begin{tabular}{llclcccc}
		\toprule
		\textbf{id} & \textbf{name}		& \textbf{gender}  & \textbf{age}	 & \textbf{race} 	& \textbf{zip code} 	& \textbf{phone} & \makecell[c]{\textbf{high} \textbf{expenditure}}			\\
		\midrule
		$t_1$ 		&	Darin Brust			& M   		& 25	& W    			& 01004    			& 2088556597	& no	\\
		$t_2$ 		&	Rosalie Bad			& F			& 22	& W    			& 01005    			& 				& no	\\
		$t_3$ 		&	Kristine Hilyard	& F			& 50	& W    			& 01004    			& 2766465009	& yes	\\
		$t_4$ 		&	Chloe Ayer			& F			& 22	& A    			&     				& 7874891021	& yes	\\
		$t_5$ 		&	Julietta Mchugh		& F			& 51	& W    			& 01009    			& 9042899033	& yes	\\
		$t_6$ 		&	Doria Ely			& F			& 32	& A    			& 01101    			& 				& yes	\\
		$t_7$ 		&	Kristan Whidden		& F			& 25	& W    			& 01101    			& 4047747803	& no	\\
		$t_8$ 		&	Rene Strelow		& M   		& 35	& W    			& 01101    			& 6162127741	& yes	\\
		$t_9$ 		&	Arial Brent			& M  		& 45	& W    			& 01102    			& 4089065769	& yes	\\
		\bottomrule
	\end{tabular}
	}
	\vspace{-2mm}
	\caption{A sample dataset \pplpass with $9$ entities. A logistic
regression classifier trained over this dataset does not discriminate against
any specific race or gender, and, thus, is fair (Example~\ref{ex:one}).}
  	\vspace{-2mm}
	\label{fig:passdb}
\end{figure}

\section{Intervention Algorithms}\label{sec:algorithm}
%

\looseness-1 We now describe our intervention algorithms to
explain the mismatch between a dataset and a system malfunctioning on that
dataset. Our algorithms consider a failing and a passing dataset
as input and report a \emph{collection} of \pvt{} triplets (or simply \pvt{}s) as the explanation
(cause and fix) of the observed mismatch. To this end, we first identify a set
of \emph{discriminative} \pvt{}s---whose profiles take different
values in the failing and passing datasets---as potential explanation units,
and then intervene on the failing dataset to alter the profiles and observe
change in system malfunction.
We develop two approaches that differ in terms of the number of \pvt{}s
considered simultaneously during an intervention. \systemgr is a greedy
approach that considers only one \pvt at a time. However, in worst case, the
number of interventions required by \systemgr is linear in number of
discriminative \pvt{}s.
Therefore, we propose a second algorithm \systemgp, built on the group-testing
paradigm, that considers multiple \pvt{}s to reduce the number of
interventions, where the number of required interventions is
\emph{logarithmically} dependent on the number of discriminative \pvt{}s. We
start with an example scenario to demonstrate how \systemgr works and then
proceed to describe our algorithms.

\subsection{Example Scenario}%
Consider the task of predicting the attribute \texttt{high\_expenditure} to
determine if a customer should get a discount (Example~\ref{ex:one}). The
system calculates bias of the trained classifier against the unprivileged
groups (measured using disparate impact~\cite{ibmaif360}) as its malfunction
score. We seek the causes of mismatch between this prediction pipeline and
\pplfail (Figure~\ref{fig:faildb}), for which the pipeline fails with a
malfunction score of $0.75$. We assume the knowledge of \pplpass
(Figure~\ref{fig:passdb}), for which the malfunction score is $0.15$. The goal
is to identify a minimal set of \pvt{}s whose transformation functions bring
down the malfunction score of \pplfail below $0.20$.

\smallskip

(Step 1) The first goal is to identify the profiles whose parameters differ
between \pplfail and \pplpass. To do so, \systemgr identifies the exhaustive
set of \pvt{}s over \pplpass and \pplfail and discards the identical ones
(\pvt{}s with identical profile-parameter values). We call the \pvt{}s of the
passing dataset whose profile-parameter values differ over the failing dataset
\emph{discriminative} \pvt{}s. Figure~\ref{fig:discpvt} lists a few profiles of
the discriminative \pvt{}s w.r.t. \pplpass and \pplfail.

\begin{figure}[t]
	\centering
	\includegraphics[width=1\columnwidth]{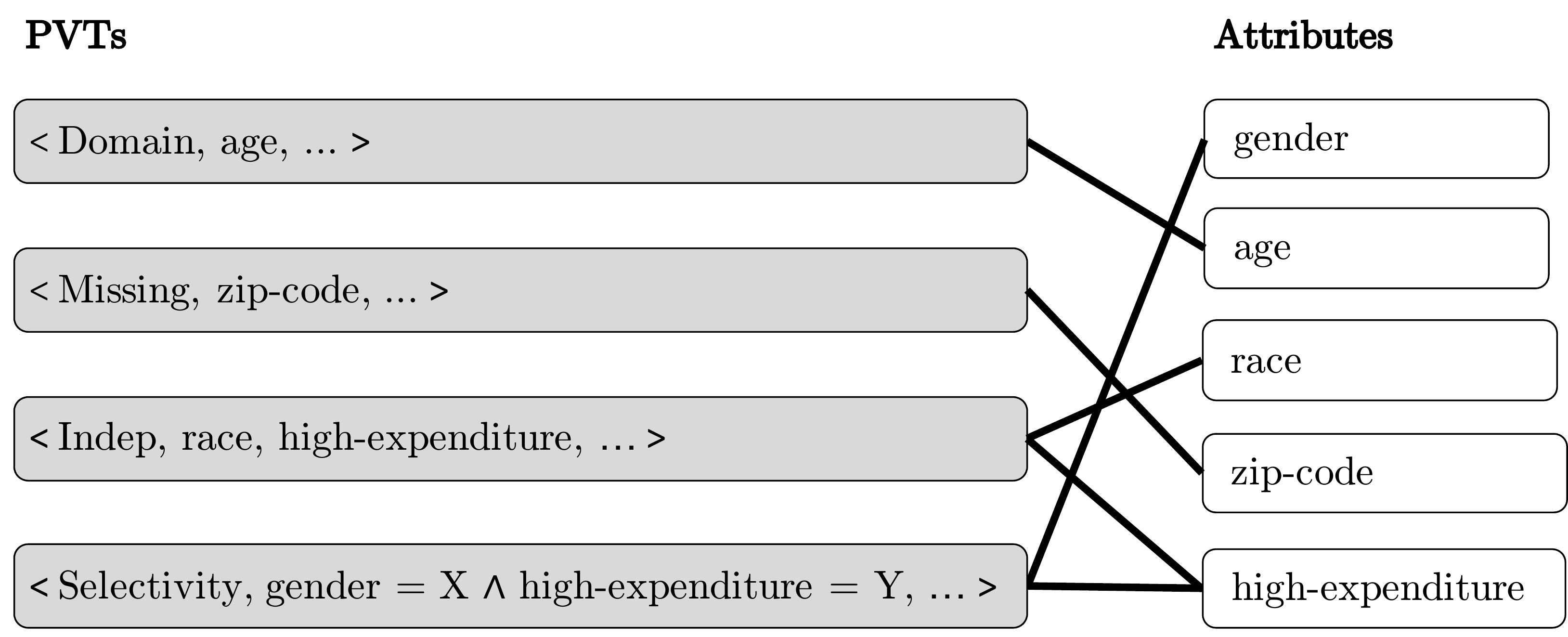}
	\vspace{-5mm}
	\caption{\pvt-attribute graph. The attribute
	\texttt{high\_expenditure} is associated
	with two discriminative \pvt{}s. For ease of exposition, we only show profile within a \pvt to denote the entire \pvt.
	}
	\vspace{-5mm}
	\label{fig:dependency}
\end{figure}

(Step 2) Next, \systemgr ranks the set of discriminative \pvt{}s based on their
likelihood of offering an explanation of the malfunction. Our intuition
here is that if an attribute $A$ is related to the malfunction, then many
\pvt{}s containing $A$ in their profiles would differ between \pplfail and
\pplpass. Additionally, altering $A$ with respect to one \pvt is likely to
automatically ``fix'' other \pvt{}s associated with $A$.\footnote{Altering
values of $A$ w.r.t. a \pvt may also increase violation w.r.t. some other
\pvt{}s. However, for ease of exposition, we omit such issues in this example
and provide a detailed discussion on such issues in the appendix.} Based on this intuition, \systemgr constructs a
\emph{bipartite graph}, called \pvt-attribute graph, with discriminative \pvt{}s on one side and data
attributes on
the other side (Figure~\ref{fig:dependency}).
In this graph, a \pvt $X$ is connected to an attribute $A$ if $X_P$ is defined
over $A$.
In the bipartite graph, the degree of an attribute $A$ captures the number of
discriminative \pvt{}s associated with $A$. During intervention, \systemgr
prioritizes \pvt{}s associated with a high-degree attributes. For instance,
since $\mathtt{high\_expenditure}$ has the highest degree in
Figure~\ref{fig:dependency}, \pvt{}s associated with
it are considered for intervention before others.

\looseness-1 (Step 3) \systemgr further ranks the subset of the discriminative
\pvt{}s that are connected to the highest-degree attributes in the \pvt-attribute
graph based on their \emph{benefit score}. Benefit score of a \pvt $X$ encodes
the likelihood of reducing system malfunction when the failing dataset is
altered using $X_T$. The benefit score of $X$ is estimated from (1)~the
violation score that the failing dataset incurs w.r.t. $X_V$, and (2)~the
number of tuples in the failing dataset that are altered by $X_T$. For example,
to break the dependence between \texttt{high\_expenduture} and \texttt{race},
the transformation corresponding to $\correlation$ modifies five tuples in
\pplfail by perturbing (adding noise to) \texttt{high\_expenditure}. In
contrast, the transformation for $\missing$ needs to change only one value
($t_6$ or $t_{10}$). Since more tuples are affected by the former, it has
higher likelihood of reducing the malfunction score. The intuition behind this
is that if a transformation alters more tuples in the failing dataset, the more
likely it is
to reduce the malfunction score. 
This holds particularly in
applications where the system optimizes aggregated
statistics such as accuracy, recall, F-score, etc. 

\begin{figure}
	\centering
	\resizebox{1\columnwidth}{!}{
	\setlength{\tabcolsep}{1mm}
	\begin{tabular}{llc|c||}
		\toprule
\pplpass	& \pplfail\\
		\midrule
\hl{$\langle \domain, \mathtt{age}, [22, 51] \rangle$}&	\hl{$\langle \domain, \mathtt{age}, [20, 60] \rangle$}\\
\hline
\hl{$\langle \missing, \mathtt{zip\_code}, 0.11 \rangle$}& \hl{$\langle \missing, \mathtt{zip\_code}, 0.2 \rangle$}\\\hline
 \hl{$\langle \correlation,\mathtt{race}, \mathtt{high\_expenditure}, 0.04 \rangle$}& \hl{$\langle \correlation,\mathtt{race}, \mathtt{high\_expenditure},0.67 \rangle$}\\\hline
 \hl{$\langle \selectivity, \mathtt{gender} = \text{F}$} &\hl{$\langle \selectivity, \mathtt{gender} = \text{F}$} \\
\hl{$ \wedge \mathtt{high\_expenditure} = \text{yes}, 0.44 \rangle$}&\hl{$\wedge \mathtt{high\_expenditure} = \text{yes} , 0.1 \rangle$}\\
		\bottomrule
	\end{tabular}
	}
	\vspace{-3mm} 
	\caption{A list of \pvt{}s that discriminate \pplpass
(Figure~\ref{fig:passdb}) and \pplfail (Figure~\ref{fig:faildb}) based on the
scenario of Example~\ref{ex:one} . We omit the violation and transformation
functions for ease of exposition.}
	\vspace{-5mm}
	\label{fig:discpvt}
\end{figure}

(Step 4) \systemgr starts intervening on \pplfail using the transformation of
the \pvt corresponding to the profile \hl{$\langle \correlation,\mathtt{race},
\mathtt{high\_expenditure}, 0.04 \rangle$} as its transformation offers the
most likely fix. Then, it evaluates the malfunction of the system over the
altered version of \pplfail. Breaking the dependence between
\texttt{high\_expenditure} and \texttt{race} helps reduce bias in the trained
classifier, and, thus, we observe a malfunction score of $0.35$ w.r.t. the
altered dataset. This exposes the first explanation of malfunction.

(Step 5) \systemgr then removes the processed \pvt (\correlation) from the
\pvt-attribute graph, updates the graph according to the altered dataset, and
re-iterates steps 2--4. Now the \pvt corresponding to the profile
$\selectivity$ is considered for intervention as it has the highest benefit
score. To do so, \systemgr over-samples tuples corresponding to female customers
with $\mathtt{high\_expenditure} = yes$. This time, \systemgr intervenes on the
transformed dataset obtained from the previous step. After this transformation,
bias of the learned classifier further reduces and the malfunction score falls
below the required threshold. Therefore, with these two interventions,
\systemgr is able to expose two issues that caused undesirable behavior of the
prediction model trained on \pplfail.

\looseness-1 (Step 6) \systemgr identifies an initial explanation over two
\pvt{}s: \correlation and \selectivity. However, to verify whether it is a
minimal, \systemgr tries to drop from it one \pvt at a time to obtain a proper
subset of the initial explanation that is also an explanation. This procedure
guarantees that the explanation only consists of \pvt{}s that are
\emph{necessary}, and, thus, is minimal. In this case, both $\correlation$ and
$\selectivity$ are necessary, and, thus, are part of the minimal explanation.
\systemgr finally reports the following as a minimal explanation of the
malfunction, where failure to satisfy the profiles is the cause and the
transformations indicate fix (violation and transformation functions are
omitted).
\begin{align*}
	\{ &\langle \correlation,\mathtt{race}, \mathtt{high\_expenditure}, 0.04
\rangle, \\
&\langle \selectivity, \mathtt{gen\-der} = \text{F} \wedge
\mathtt{high\_expenditure} = \text{yes}, 0.44 \rangle\}
\end{align*}

\subsection{Assumptions and Observations\label{sec:observations}} We now
proceed to describe our intervention algorithms more formally. We first state
our assumptions and then proceed to present our observations that lead to the
development of our algorithms.

\subsubsection*{Assumptions} \system makes the following assumptions:

(A1) The ground-truth explanation of malfunction is captured by at least one of
the discriminative \pvt{}s. This assumption is prevalent in software-debugging
literature where program predicates are assumed to be expressive enough to
capture the root causes~\cite{aid, liblit2005scalable}.


(A2) If the fix corresponds to a composition of transformations, then the
malfunction score achieved after applying the composition of transformations is
less than the malfunction score achieved after applying any of the
constituents, and all these scores are less than the malfunction score of the
original dataset. For example, consider two discriminative \pvt{}s $X$ and $Y$
and a failing dataset $D_{\mathit{fail}}$. Our assumption is that if $\{X, Y\}$
corresponds to a minimal explanation, then $m_S((Y_\T \circ
X_\T)(D_{\mathit{fail}})) < m_S(X_\T(D_{\mathit{fail}})) <
m_S(D_{\mathit{fail}})$ and $m_S((Y_\T \circ X_\T)(D_{\mathit{fail}})) <
m_S(Y_\T(D_{\mathit{fail}})) < m_S(D_{\mathit{fail}})$.
Intuitively, this assumption states that $X$ and $Y$ have consistent
(independent) effect on reducing the malfunction score, regardless of whether
they are intervened together or individually. If this assumption does not hold,
\system can still work with additional knowledge about multiple failing and
passing datasets. More details are in the appendix.

\subsubsection*{Observations} We make the following
observations:

(O1) If the ground-truth explanation of malfunction corresponds to an
attribute, then multiple \pvt{}s that involve the same attribute are likely to
differ across the passing and failing datasets. This observation motivates us
to prioritize interventions based on \pvt{}s that are associated with
high-degree attributes in the \pvt-attribute graph. Additionally, intervening
on the data based on one such \pvt is likely to result in an automatic ``fix''
of other \pvt{}s connecting via the high-degree attribute. For example, adding
noise to \texttt{high\_expenditure} in Example~\ref{ex:one} breaks its
dependence with not only \texttt{race} but also with other attributes.

(O2) The \pvt for which the failing dataset incurs higher violation score is
more likely to be a potential explanation of malfunction.

(O3) A transformation function that affects a large number of data tuples is
 likely to result in a higher reduction in the malfunction score, after the
transformation is applied.

\smallskip

\noindent \textbf{\pvt-attribute graph.} \system leverages observation O1 by
constructing a bipartite graph ($G_{PA}$), \emph{called \pvt-attribute graph,} with
all attributes $A \in \mathcal{R}(D)$ as nodes on one side and all
discriminative \pvt{}s $X\in \P$ on the other side. An attribute $A$ is
connected to a \pvt $X$ if and only if $X_P$ has $A$ as one of its parameters.
E.g., Figure~\ref{fig:dependency} shows the \pvt-attribute graph w.r.t.
\pplfail and \pplpass (Example~\ref{ex:one}). In this graph, the \pvt
corresponding to \pl{$\langle \correlation, \mathtt{race},
\mathtt{high\_expenditure}\rangle$} is connected to two attributes,
\texttt{race} and \texttt{high\_expenditure}. Intuitively, this graph captures
the dependence relationship between \pvt{}s and attributes, where an
intervention with respect to a \pvt $X$ modifies an attribute $A$ connected to
it. If this intervention reduces the malfunction score then it could possibly
fix other \pvt{}s that are connected to $A$.

\setlength{\textfloatsep}{5pt}
\setlength{\floatsep}{2pt}
 
\begin{algorithm}[t]
\KwIn{Failing dataset $D_{\mathit{fail}}$, passing dataset $D_{\mathit{pass}}$, malfunction score threshold $\tau$}
\KwOut{A minimal explanation set of \pvt{}s $\E$}

$\P_f \leftarrow \textsc{Discover-PVT}(D_{\mathit{fail}})$ \label{lineone}\\
$\P_p \leftarrow \textsc{Discover-PVT}(D_{\mathit{pass}})$ \label{linetwo}\\
$\P_\cap \leftarrow \P_f\cap \P_p$ \label{linethree} \tcc*[f]{Common \pvt{}s}\\
$\P \leftarrow \P_p \setminus \P_\cap$ \label{linefour} \tcc*[f]{Discriminative \pvt{}s}\\
$G_{PA} (V_G, E_G) \leftarrow \textsc{Construct-\pvt-Attr-Graph}(\P, D_{\mathit{fail}})$ \label{linefive}\\
$B \leftarrow \textsc{Calculate-Benefit-Score}(\P, G_{PA}, D_{\mathit{fail}})$\label{linesix} \\
$\E \leftarrow \emptyset$ \tcc*[f]{Initialize minimal explanation set to be empty}\label{lineseven}\\
$D \leftarrow D_{\mathit{fail}}$ \tcc*[f]{Initialize dataset to the failing dataset}\label{lineeight}\\
\While{$m_S(D) > \tau$\label{linenine}}
{
	$\P_{\mathtt{hda}} = \{ X {\in} \P | (X, A) {\in} E_G {\wedge} A {=} \argmax_{A \in \mathcal{R}(D)} deg_{G}(A)\}\!\!$ \label{lineten} 
	\tcc*[f]{\pvt{}s adjacent to high-degree attributes in $G_{PA}$}\\
	$X = \argmax_{X \in \P_{\mathtt{hda}}} B(X)$ \label{line11} \tcc*[f]{Highest-benefit \pvt}\\
	$\Delta \leftarrow m_S(D)- m_S(X_\T(D))$ \label{line12} \tcc*[f]{Malfunction reduction}\\
	$G_{PA}\leftarrow G_{PA}.\textsc{Remove}(X)\label{line13}$ \tcc*[f]{Update $G_{PA}$} \\
	\If(\tcc*[f]{Reduces malfunction}){$\Delta > 0$\label{line14}}
	{
		 $D \leftarrow X_\T(D)$ \tcc*[f]{Apply transformation}\\
		  $G_{PA}.\textsc{Update}(D)$ \tcc*[f]{Update the \pvt-attribute graph }\\
		 $B.\textsc{Update}(D)$ \tcc*[f]{Update benefit scores}\\
		 $\E \leftarrow \E \cup \{ X\}$ \tcc*[f]{Add $P$ to explanation set}\\
		 $\P \leftarrow \P \setminus \{ X\}$\label{line18} \tcc*[f]{Remove $P$ from the candidates}\\
	}
}
$\E$ = \textsc{Make-Minimal}($\E$)\label{line19} \tcc*[f]{Obtain minimality of $\E$}\\
\Return $\E$ \tcc*[f]{$\E$ is a minimal explanation}

\caption{\systemgr (greedy)}
\label{alg:datadebuggreedy} 
\end{algorithm}

\smallskip \noindent \textbf{Benefit score calculation.} \system uses the
aforementioned observations to compute a benefit score for each \pvt to model
their likelihood of reducing system malfunction if the corresponding
transformation is used to modify the failing dataset $D_{\mathit{fail}}$.
Intuitively, it assigns a high score to a \pvt with a high violation score (O2)
and if the corresponding transformation function modifies a large number of
tuples in the dataset (O3). Formally, the benefit score of a \pvt $X$ is
defined as the product of violation score of $D_{\mathit{fail}}$ w.r.t. $X_V$
and the ``coverage'' of $X_\T$. The coverage of $X_\T$ is defined as the
fraction of tuples that it modifies. Note that the benefit calculation
procedure acts as a proxy of the likelihood of a \pvt to offer an explanation,
without actually applying any intervention.

\subsection{Greedy Approach} Algorithm~\ref{alg:datadebuggreedy} presents the
pseudocode of our greedy technique \systemgr, which takes a passing dataset
$D_{\mathit{pass}}$ and a failing dataset $D_{\mathit{fail}}$ as input and
returns the set of \pvt{}s that corresponds to a minimal explanation of system
malfunction.

\begin{description}
\item[\algoexp{Lines~\ref{lineone}-\ref{linetwo}}] Identify two sets of \pvt{}s
$\P_f$ and $\P_p$ satisfied by $D_{\mathit{fail}}$ and $D_{\mathit{pass}}$, respectively.

\item[\algoexp{Lines~\ref{linethree}-\ref{linefour}}] Discard the \pvt{}s
$\P_f \cap \P_p$ from $\P_p$ and consider the remaining discriminative ones
$\P\equiv \P_p\setminus \P_f$ as candidates for potential explanation of system malfunction.

\item[\algoexp{Line~\ref{linefive}}] Compute the \pvt-attribute graph $G_{PA}$,
where the candidate \pvt{}s $\P$ correspond to nodes on one side and the data
attributes correspond to nodes on the other side.

\item[\algoexp{Line~\ref{linesix}}] Calculate the benefit score of each
discriminative \pvt $X \in \P$ w.r.t. $D_{\mathit{fail}}$. This procedure
relies on the violation score using the violation function of the \pvt and the
coverage of the corresponding transformation function over $D_{\mathit{fail}}$.

\item[\algoexp{Line~\ref{lineseven}-\ref{lineeight}}] Initialize the solution set
$\mathcal{X}^*$ to $\emptyset$ and the dataset to perform intervention on $D$
to the failing dataset $\D_{\mathit{fail}}$. In subsequent steps, $\mathcal{X}^*$ will
converge to a minimal explanation set and $D$ will be transformed to a dataset for
which the system passes.

\item[\algoexp{Line~\ref{linenine}}] Iterate over the candidate \pvt{}s $\P$
until the dataset $D$ (which is being transformed iteratively) incurs an
acceptable violation score (less than the allowable
threshold $\tau$).

\item[\algoexp{Line~\ref{lineten}}] Identify the subset of \pvt{}s
$\P_{\mathtt{hda}}\subseteq \P$ such that all $X\in \P_{\mathtt{hda}}$ are
adjacent to at least one of the highest degree attributes in the current
\pvt-attribute graph (Observation O1).

\item[\algoexp{Line~\ref{line11}}] Choose the \pvt $X\in \P_{\mathtt{hda}}$
that has the maximum benefit.

\item[\algoexp{Line~\ref{line12}}] Calculate the reduction in malfunction score
if the dataset $D$ is transformed according to the transformation $X_\T$.

\item[\algoexp{Line~\ref{line13}}] Remove $X$ from $G_{PA}$ as it has been
explored. \item[\algoexp{Lines~\ref{line14}-\ref{line18}}] If the malfunction
score reduces over $X_\T(D)$, then $X$ is added to the solution set $\E$, and
$D$ is updated to $X_\T(D)$, which is then used to update the \pvt-attribute
graph and benefit of each \pvt. The update procedure recalculates the benefit
scores of all \pvt{}s that are connected to the attributes adjacent to $X$ in
$G_{PA}$.

\item[\algoexp{Line~\ref{line19}}] Post-process the set $\E$ to identify a
minimal subset that ensure that malfunction score remains less than the
threshold $\tau$. This procedure iteratively removes one \pvt at a time (say
$X$) from $\E$ and recalculates the malfunction score over the failing dataset
$D_{\mathit{fail}}$ transformed according to the transformation functions of
the \pvt{}s in the set $\P'= \E\setminus \{X\}$. If the transformed dataset
incurs a violation score less than $\tau$ then $\E$ is replaced with $\P'$.
\end{description}

\subsection{Group-testing-based Approach\label{sec:grouptest}} We now present
our second algorithm \systemgp, which performs group interventions to identify
the minimal explanation that exposes the mismatch between a dataset and a
system. The group intervention methodology is applicable under the following
assumption along with assumptions $A_1$ and $A_2$
(Section~\ref{sec:observations}).

(A3) The
malfunction score incurred after applying a composition of transformations is less than the
malfunction score incurred by the the original dataset if and only if at least one of the
constituent transformations reduces the malfunction
score.
For two \pvt{}s $X$ and $Y$, $m_S((X_\T \circ Y_\T)(D_{\mathit{fail}})) <
m_S(D_{\mathit{fail}})$, iff $m_S(X_\T (D_{\mathit{fail}})) <
m_S(D_{\mathit{fail}})$ or $m_S(Y_\T (D_{\mathit{fail}})) <
m_S(D_{\mathit{fail}})$. Note that this assumption is crucial to consider group
interventions and is prevalent in the group-testing
literature~\cite{du2000combinatorial}.
 
\looseness-1 \systemgp follows the classical adaptive group testing (GT)
paradigm~\cite{du2000combinatorial} for interventions. To this end, it
iteratively partitions the set of discriminative \pvt{}s into two ``almost''
equal subsets (when the number of discriminative \pvt{}s is odd, then the size
of the two partitions will differ by one). During each iteration, all \pvt{}s
in a partition are considered for intervention \textit{together} (group
intervention) to evaluate the change in malfunction score. If a partition does
not help reduce the malfunction score, all \pvt{}s within that partition are
discarded. While traditional GT techniques~\cite{du2000combinatorial} would use
a random partitioning of the \pvt{}s, \systemgp can leverage the dependencies
among \pvt{}s (inferred from the \pvt-attribute graph) to achieve more
effective partitioning. Intuitively, it is beneficial to assign all \pvt{}s
whose transformations operate on the same attribute to the same partition,
which is likely to enable aggressive pruning of spurious \pvt{}s that do not
reduce malfunction.

\looseness-1 \systemgr captures the dependencies among \pvt{}s by constructing
a \pvt-dependency graph $G_{PD}$. Two \pvt{}s $U$ and $V$ are connected by an
edge in $G_{PD}$ if they are connected via some attribute in $G_{PA}$. $G_{PD}$
is equivalent to $G_{PA}^2$ (transitive closure of $G_{PA}$), restricted to
\pvt nodes (excluding the attribute nodes). This ensures that \pvt{}s that are
associated via some attribute in $G_{PA}$ are connected in $G_{PD}$. \systemgr
partitions $G_{PD}$ such that the number of connections (edges) between \pvt{}s
that fall in different partitions are minimized.
More formally, we aim to construct two ``almost'' equal-sized partitions of
$\P$ such that the number of edges between \pvt{}s from different partitions
are minimized, which maps to the problem of finding the \textit{minimum
bisection} of a graph~\cite{garey1974some}. The minimum bisection problem is
NP-hard~\cite{garey1974some} and approximate algorithms
exist~\cite{fernandes2018minimum,fellows2012local}. In this work, we use the
\textit{local search algorithm}~\cite{fellows2012local} (details are in the
appendix).

We proceed to demonstrate the benefit of using \systemgp as opposed to traditional GT with the following
example.

\begin{algorithm}[t]
\KwIn{Failing dataset $D_{\mathit{fail}}$, passing dataset $D_{\mathit{pass}}$, malfunction score threshold $\tau$}
\KwOut{A minimal explanation set of \pvt{}s $\E$}

$\P_f \leftarrow \textsc{Discover-PVT}(D_{\mathit{fail}})$\\
$\P_p \leftarrow \textsc{Discover-PVT}(D_{\mathit{pass}})$ \\
$\P_\cap \leftarrow \P_f\cap \P_p$ \tcc*[f]{Common \pvt{}s}\\
$\P \leftarrow \P_p \setminus \P_\cap$ \tcc*[f]{Discriminative \pvt{}s}\\
$G_{PA} (V_G, E_G) \leftarrow \textsc{Construct-\pvt-attr-Graph}(\P, D_{\mathit{fail}})$ \\
$D, \E \leftarrow  \textsc{Group-Test}(\P,D_{\mathit{fail}}, G_{PA}^2, \tau)$ \tcc*[f]{Obtain an exp.}\\
$\E$ = \textsc{Make-Minimal}($\E$) \tcc*[f]{Obtain minimality of $\E$}\\
\Return $\E$ \tcc*[f]{$\E$ is a minimal cause}

\caption{\systemgp (group-testing-based)}
\label{alg:datadebuggrptest} 
\end{algorithm}
 
\begin{algorithm}[t]
 \KwIn{Candidate \pvt $\P$, dataset $D$, \pvt-dependency graph~$G_{PD}$, malfunction score threshold $\tau$}
 \KwOut{A transformed dataset $D'$ and an explanation set of \pvt{s} $\E$ 
 }


 $\E \leftarrow \emptyset$ \label{gp:lineone}\tcc*[f]{Initialize explanation set to be empty}\\

 \If(\label{gp:linetwo}\tcc*[f]{Only a single \pvt is candidate}){$|\P| = 1$}
 {
 	\Return $\T_{\P}(D), \P$\label{gp:linethree}
 }
 $\P^1,\P^2 \leftarrow  \textsc{Get-Min-Bisection} (G_{PD},\P)$\label{gp:linefour} \tcc*[f]{Partition $\P$ }\\
 $\mathcal{M} \leftarrow m_S(D)$\label{gp:linefive}\tcc*[f]{Initial malfunction score}\\
 $\Delta_1 \leftarrow \mathcal{M} - m_S(\P^1_\T(D))$\label{gp:linesix} \tcc*[f]{Malfunction reduction by $\P^1_\T$}\\

 \If(\tcc*[f]{${\P^1}$ alone is insufficient}){$\mathcal{M} - \Delta_1 > \tau$\label{gp:lineseven}}
 {
 	$\Delta_2 \leftarrow \mathcal{M} {-} m_S(\P^2_\T(D))$ \label{gp:lineeight}\tcc*[f]{Malfunction reduction by $\P^2_\T$}\\
 }

 \If{$(\mathcal{M} - \Delta_1 \le \tau)$ OR $(\Delta_1 > 0$ AND $\mathcal{M} - \Delta_2 > \tau)$\label{gp:linenine}}
 {
 	\tcc*[f]{\mbox{${\P_1}$ is sufficient OR ${\P_1}$ helps AND ${\P_2}$ is insufficient}}\\
 	$D, \mathcal{X}' \leftarrow  \textsc{Group-Test}(\P_1,D, G_{PD})$\\
 	$\E = \E \cup  \mathcal{X}'$ \tcc*[f]{Augment explanation set}

 	\If(\tcc*[f]{Malfunction is acceptable}){$\mathcal{M} - \Delta_1 \le \tau$}
 	{
 		\Return $D, \E$ \tcc*[f]{No need to check $\P_2$\label{gp:line13}}
 	}
 }

 \If(\tcc*[f]{$\T_{\P_2}$ reduces malfunction}){$\Delta_2 > 0$\label{gp:line14}}
 {
 	$D, \mathcal{X}' \leftarrow  \textsc{Group-Test}(\P_2,D, G_{PD})$\\
 	$\E = \E \cup  \mathcal{X}'$ \label{gp:line16}\tcc*[f]{Augment explanation set}
 }
 \Return $D, \E$

 \caption{\textsc{Group-Test}}
 \label{alg:grptest} 
 \end{algorithm}

\setlength{\textfloatsep}{10pt}

\begin{figure}
    \centering
    \begin{subfigure}[t]{0.24\columnwidth}
		\includegraphics[width=\columnwidth]{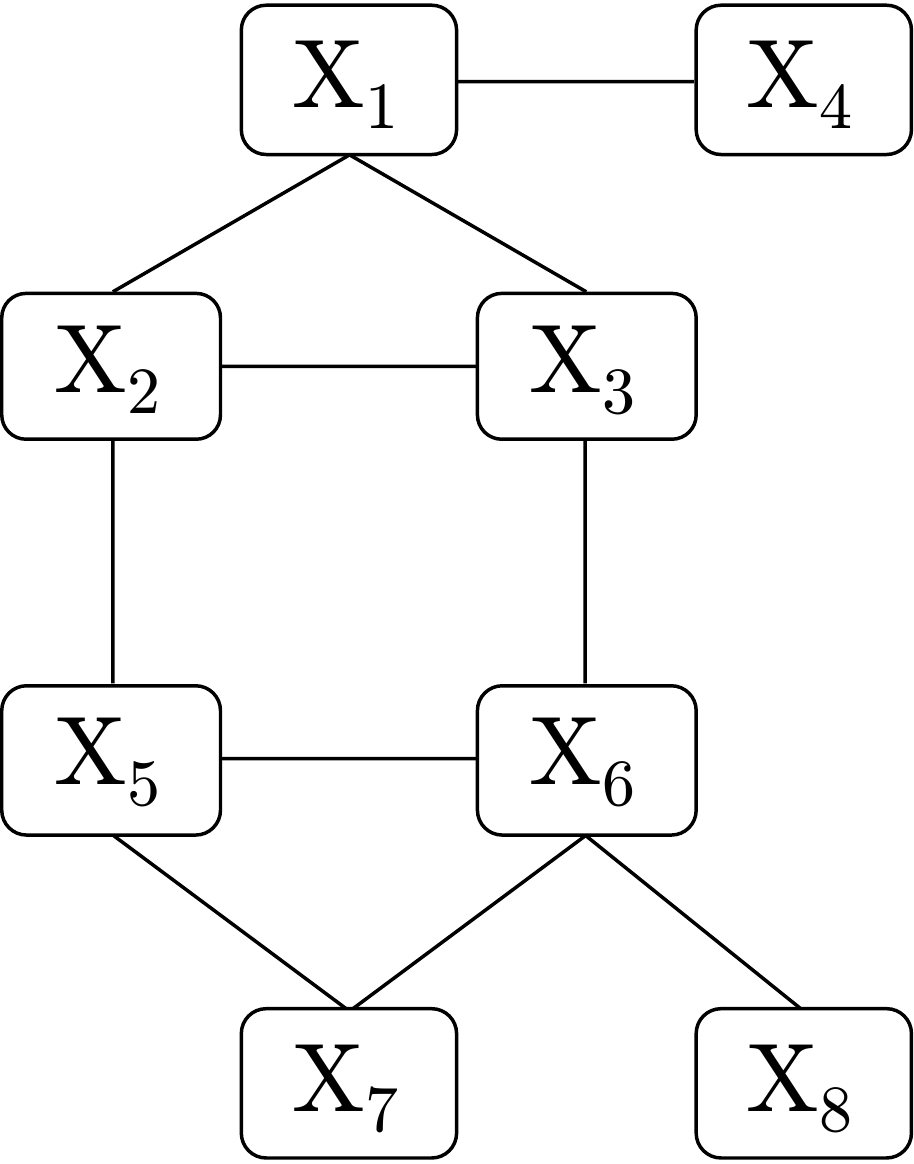}
		\caption{Dependency graph $G_{PD}$}
	\end{subfigure}
	\hspace{2mm}
    \begin{subfigure}[t]{0.69\columnwidth}
		\includegraphics[width=\columnwidth]{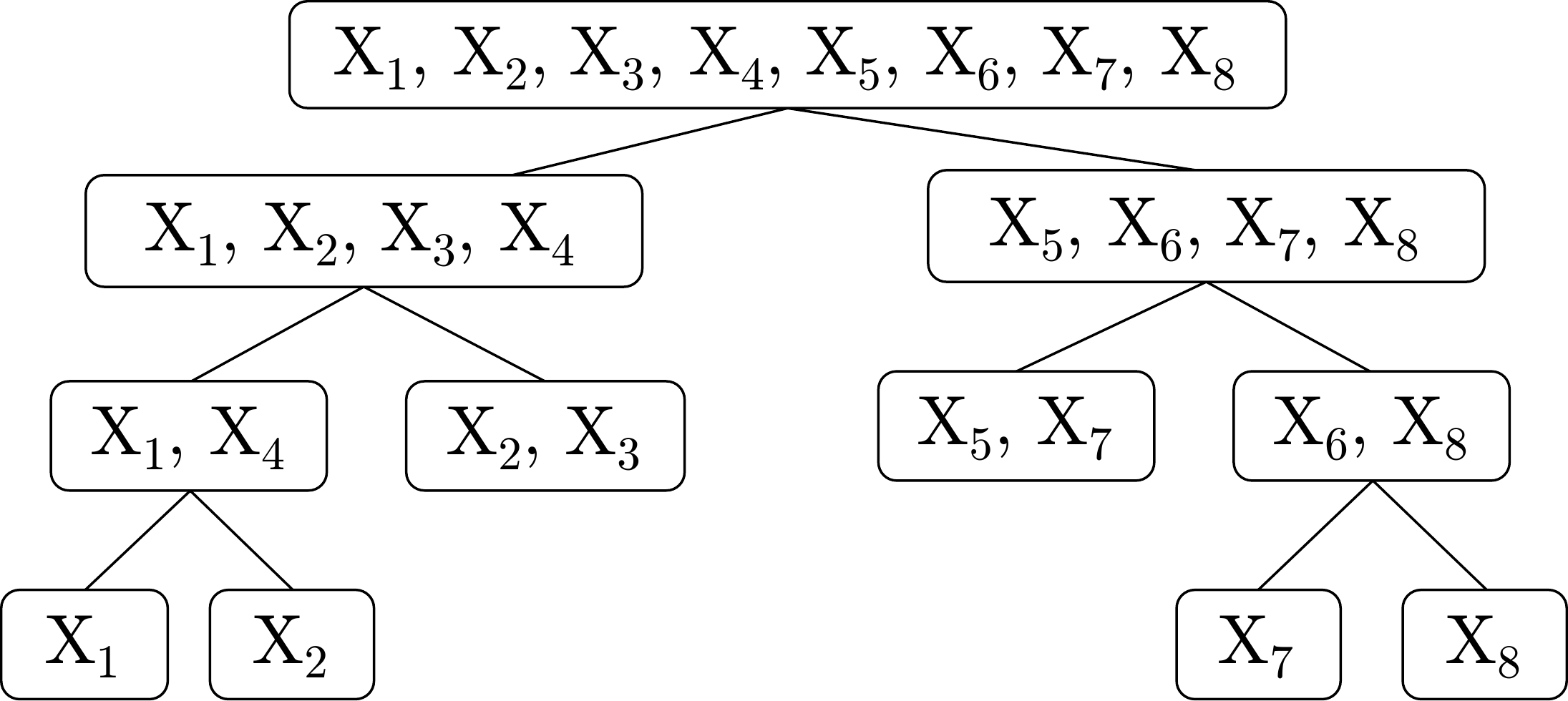}
		\caption{Execution of \systemgp}
    \end{subfigure}
    \vspace{-2.5mm}
    \begin{subfigure}[t]{0.67\columnwidth}
		\includegraphics[width=\columnwidth]{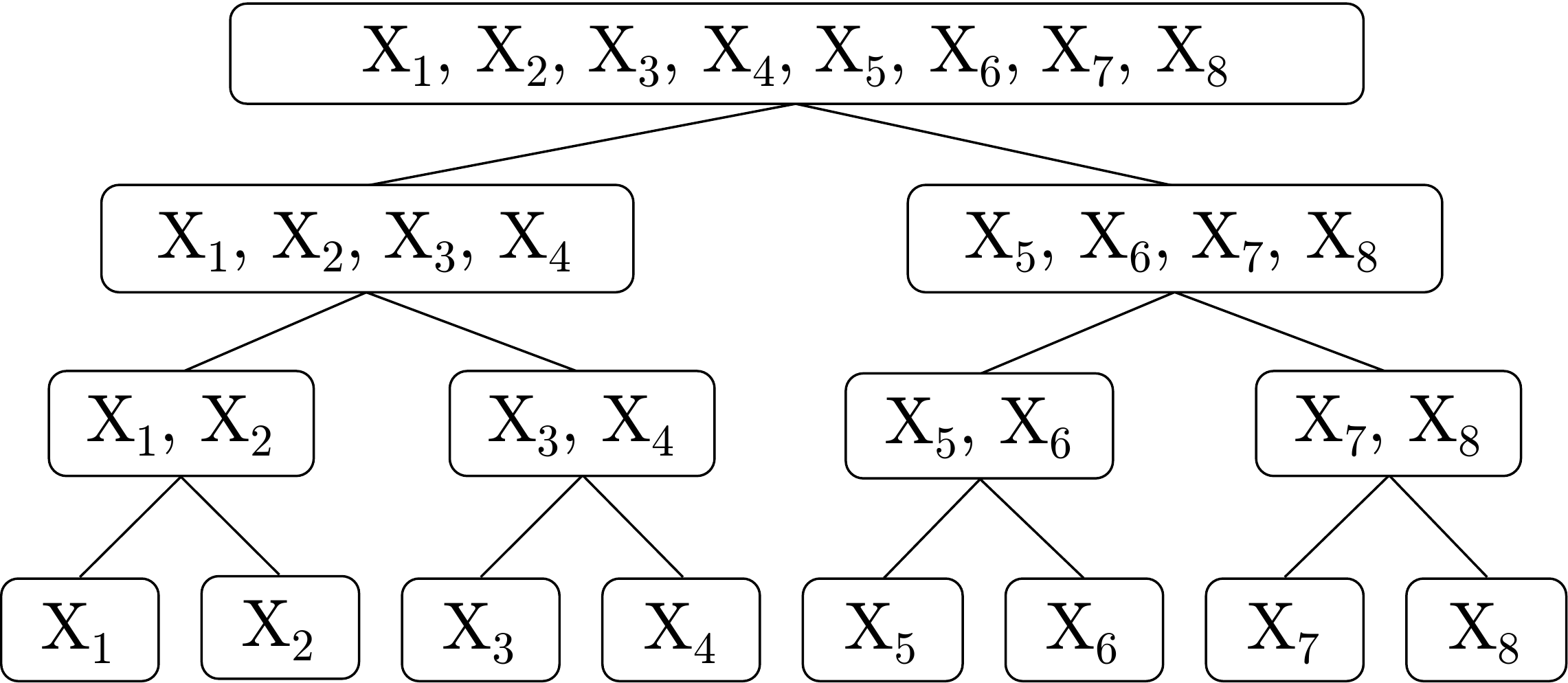}
		\caption{Execution of traditional group testing algorithm}
		\label{fig:grprandom}
    \end{subfigure}
	\vspace{-2mm}
    \caption{Comparison between \systemgp and adaptive group testing on a toy example.}
	\vspace{-3.5mm}
    \label{fig:grptestcomparison}
\end{figure}

\begin{example}
	
\looseness-1 Consider a set of $\;8$ \pvt{}s $\P=\{X_1,\ldots,X_{8}\}$ where the ground-truth (minimal) explanation is either $\{X_1,X_6\}$ or
$\{X_4,X_8\}$ (disjunction). An example of steps for a  traditional adaptive GT approach is shown in
Figure~\ref{fig:grprandom}. In this case, it requires a total of $14$
interventions. Note that adaptive GT is a randomized algorithm and this example
demonstrates one such execution. However, we observed similar results for other
instances. In contrast to adaptive GT, \systemgp constructs a min-bisection of
the graph during each iteration: it does not partition $\{X_2,X_3\}$ and
$\{X_5,X_7\}$ as none of these \pvt{}s help reduce the malfunction. Therefore,
it requires only $10$ interventions.
\end{example}

 Algorithm~\ref{alg:datadebuggrptest} presents the \systemgp algorithm. It
starts with a set of discriminative \pvt{}s $\P$ and the \pvt-attribute graph
$G_{PA}$. All candidate \pvt{}s are then considered by \textsc{Group-Test}
subroutine to identify the explanation $\E$.

\smallskip

\paragraph{\textsc{Group-Test}.} Algorithm~\ref{alg:grptest} presents the procedure
 that takes the set of discriminative \pvt{}s $\P$,  a failing dataset $D$, \pvt-dependency graph $G_{PD}$,
 and the malfunction score threshold $\tau$ as input. It returns a transformed (fixed) dataset and an explanation.
\begin{description}
\item[\algoexp{Lines~\ref{gp:lineone}}]  Initialize the solution set $\E$ to $\emptyset$.
\item[\algoexp{Lines~\ref{gp:linetwo}-\ref{gp:linethree}}] Return the candidate \pvt set $\P$ 
if its cardinality is 1.
\item[\algoexp{Lines~\ref{gp:linefour}}]  Partition $\P$ into $\P_1$ and $\P_2$ using min-bisection of the \pvt-dependency graph $G_{PD}$.
\item[\algoexp{Lines~\ref{gp:linefive}}] Calculate the malfunction score of the input dataset.
\item[\algoexp{Lines~\ref{gp:linesix}}] Calculate the reduction in malfunction score $\Delta_1$ if $D$ is intervened w.r.t. all \pvt{}s $\P_1$.
\item[\algoexp{Lines~\ref{gp:lineseven}-\ref{gp:lineeight}}] If the malfunction exceeds $\tau$
even after intervening on $D$ w.r.t. all \pvt{}s
in $\P_1$ then try out $\P_2$: calculate the reduction in malfunction score $\Delta_2$ if $D$ is intervened w.r.t. all \pvt{}s 
in $\P_2$.
\item[\algoexp{Lines~\ref{gp:linenine}-\ref{gp:line13}}] Recursively call 
\textsc{Group-Test} on the partition $\P_1$ if one of the
following conditions hold:
(1) Intervening on $D$ w.r.t. all \pvt{}s in $\P_1$ reduces the  malfunction to be lower than $\tau$: 
the explanation over $\P_1$ is returned as the final explanation.
(2) Intervening on $D$ w.r.t. all \pvt{}s in $\P_1$ reduces the malfunction, but still remains above $\tau$, but intervening on $D$ w.r.t. all \pvt{}s in $\P_2$ brings the malfunction below $\tau$: the explanation returned by the recursive call on $\P_1$ is added to the set $\E$ and $\P_2$ is processed next.
 \item[\algoexp{Lines~\ref{gp:line14}-\ref{gp:line16}}] Recursively call 
 \textsc{Group-Test} on $\P_2$ if 
 intervening on all \pvt{}s in $\P_2$ reduces malfunction. 
 The set of \pvt{}s returned by this recursive call of the
 algorithm are added to the solution set $\E$.
 
\end{description}

\smallskip
\paragraph{Discussion on \systemgr vs. \systemgp.} \looseness-1
\systemgr intervenes by considering a single discriminative \pvt at a
time. Hence, in the worst case, it requires $O(|\P|)$ interventions where $\P$
denotes the set of discriminative \pvt{}s. Note that \systemgr requires
much fewer interventions in practice and would require $O(|\P|)$ only when any of the
mentioned observations (O1--O3) do not hold.
In contrast, \systemgp performs
group intervention by recursively partitioning the set of discriminative \pvt{}s. Thus, the maximum number of interventions required by \systemgp is
$O(t\log |\P|)$ where $t$ denotes the number of \pvt{}s that help
reduce malfunction if the corresponding profile is altered. Note that, in expectation,
\systemgp 
requires fewer interventions than \systemgr whenever
$t=o(|\P|/\log |\P|)$. \systemgp is particularly helpful when multiple \pvt{}s \textit{disjunctively} explain the malfunction.
However,
\systemgp requires an additional assumption assumption A3 (Section~\ref{sec:grouptest}).
We discuss the empirical impact of this assumption in Section~\ref{realworld}
(Cardiovascular disease prediction).
Overall, we conclude that \systemgp is beneficial for applications whenever
$t=O(|\P|/\log |\P|)$ and observations O1-O3 hold (more details are in the appendix).

\section{Experimental Evaluation}\label{sec:experiment}
%

Our experiments involving \system aim to answer the following questions:
(Q1)~Can \system \emph{correctly} identify the cause and corresponding fix of
mismatch between a system and a dataset for which the system fails? (Q2)~How
\emph{efficient} is \system compared to other alternative techniques? (Q3)~Is
\system \emph{scalable} with respect to the number of discriminating \pvt{}s?

\smallskip

\begin{figure*}[ht]
\vspace{-3mm}
\footnotesize
    \centering
    \begin{tabular}{|c|c|c|c|c|c||c|c|c|c|c|}
    \hline
    &\multicolumn{5}{c||}{\textbf{Number of Interventions}} & \multicolumn{5}{c|}{\textbf{Execution Time (seconds)}} \\
    \hline
       Application  &  \systemgr & \systemgp & \texttt{BugDoc} & \texttt{Anchor} & \texttt{GrpTest} & \systemgr & \systemgp & \texttt{BugDoc} & \texttt{Anchor} & \texttt{GrpTest}\\
       \hline\hline
       \textbf{Sentiment}& 2&3&10&303&3& 25.1& 23.4&64.6&4594.9&21.2\\
       \textbf{Income}  & 1&8&20&800&10& 11.8&12.5&20.0&195.5&10.4 \\
       \textbf{Cardiovascular}&1&NA&100&5900&NA& 7.6&NA&62.1&8602.9&NA \\\hline
    \end{tabular}
	\vspace{-3mm}
    \caption{Comparison of number of interventions and running time of \system with other baselines. NA denotes that the technique could not identify the cause of malfunction because assumption A3 did not hold.}
	\vspace{-3.5mm}
    \label{tab:comparison}
\end{figure*}

\smallskip

\paragraph{Baselines.} Since there is no prior work on modifying a dataset
according to a \pvt, 
we adapted state-of-the-art debugging and explanation techniques to incorporate profile transformations and explain the cause of system failure. We consider three baselines:
(1)~\texttt{BugDoc}~\cite{bugdoc} is a recent debugging technique that explores
different parameter configurations of the system to understand its behavior. We
adapt \texttt{BugDoc} to consider each \pvt as a parameter of the system and
interventions as the modified configurations of the pipeline.
(2)~\texttt{Anchor}~\cite{anchors} is a local explanation technique for
classifiers that explains individual predictions based on a surrogate model. We
train \texttt{Anchor} with \pvt{}s as features, and the prediction variable is
Pass/Fail where Pass (Fail) denotes the case where an input dataset incurs
malfunction below (above) $\tau$. In this technique, each intervention creates
a new data point to train the surrogate model.
(3)~\texttt{GrpTest}~\cite{du2000combinatorial} is an adaptive group testing
approach that performs group interventions to expose the mismatch between the
input dataset and the system. It is similar to \systemgp with a difference that
the recursive partitioning of \pvt{}s is performed randomly without exploiting
the \pvt-dependency graph.

\subsection{Real-world Case Studies}\label{realworld}

We design three case studies focusing on three different applications, where we
use well-known ML models~\cite{akbik2018coling, adaboost, randomforest} as
black-box systems. For all of the three case studies, we use real-world
datasets. Figure~\ref{tab:comparison} presents a summary of our evaluation
results.

\smallskip

\paragraph{Sentiment Prediction.}
The system in this study predicts sentiment of input text (reviews/tweets) and computes misclassification
rate as the malfunction score. It uses flair~\cite{akbik2018coling}, a
pre-trained classifier to predict sentiment of the input records and assumes a \texttt{target} attribute in the
input data, indicating the ground truth sentiment: A value of $1$ for the
attribute \texttt{target} indicates positive sentiment and a value of $-1$
indicates negative sentiment. We test the system over \texttt{IMDb}
dataset~\cite{imdb} ($50K$ tuples) and a \texttt{twitter}
dataset~\cite{sentiment140} (around $1.6$M tuples). The malfunction score of
the system on the \texttt{IMDb} dataset is only $0.09$ while on the
\texttt{twitter} dataset it is $1.0$.
We considered \texttt{IMDb} as the passing dataset and \texttt{twitter} as the
failing dataset and used
both \systemgr and \systemgp to explain the mismatch between the \texttt{twitter}
dataset and the system. The ground-truth cause of the malfunction is that the
\texttt{target} attribute in the \texttt{twitter} dataset uses ``4'' to denote
positive and ``0'' to denote negative sentiment~\cite{sentiment140}. 
\systemgr identifies a total of $3$ discriminative \pvt{}s between the two
datasets. One such \pvt includes the profile $\domain$ of the \texttt{target} attribute that has corresponding parameter $\mathbb{S}=\{-1,1\}$ for \texttt{IMDb} and $\mathbb{S}=\{0,4\}$
for the \texttt{twitter} dataset. \systemgr performs two
interventions and identifies that the malfunction score reduces to $0.36$ by
mapping $0\rightarrow -1$ and $4\rightarrow 1$ by intervening w.r.t.  the \pvt corresponding $\domain$, which is returned as an explanation of the malfunction.

\systemgp and \texttt{GrpTest} both require 
$3$ group interventions to explain the cause 
of system malfunction. \texttt{BugDoc} and \texttt{Anchor}
require $10$ and $303$ interventions, respectively. \texttt{Anchor} calculates
system malfunction on datasets transformed according to various local
perturbations of the \pvt{}s in the failing dataset. 

\smallskip

\paragraph{Income Prediction.}
The system in this study trains a Random Forest classifier~\cite{randomforest}
to predict the income of individuals while ensuring fairness towards
marginalized groups. The pipeline returns the normalized disparate
impact~\cite{ibmaif360} of the trained classifier w.r.t.\ the protected
attribute (\texttt{sex}), as the malfunction score. Our input data includes
census records~\cite{Dua:2019} containing demographic attributes of individuals
along with information about income, education, etc. We create two datasets
through a random selection of records, and manually add noise to one of them to
break the dependence between \texttt{target} and \texttt{sex}.
 The system has  malfunction score of $0.195$ for the passing dataset and $0.58$ for the failing dataset due to the dependence
between \texttt{target} and \texttt{sex}.
\systemgr identifies a total of $43$ discriminative \pvt{}s and constructs a \pvt-attribute graph. In this graph, the 
\texttt{target} attribute has degree $15$ while all other attributes have degree $2$.
The \pvt{}s that include \texttt{target} are then intervened in
non-increasing order of benefit. The 
transformation w.r.t. $\correlation$ \pvt on the \texttt{target} attribute breaks the dependence
between \texttt{target} and all other attributes, thereby reducing the
malfunction score to $0.32$. Therefore, \systemgr requires one intervention to explain the cause of the malfunction. 
Our group testing algorithm \systemgp and \texttt{GrpTest} require $8$ and $10$ interventions, respectively.
Note that group testing is not very useful because the datasets contain few discriminative \pvt{}s.

\texttt{BugDoc} and \texttt{Anchor} do not identify discriminative \pvt{}s explicitly and consider
all \pvt{}s ($136$ for this dataset) as candidates for intervention. \texttt{Anchor} performs $800$ local interventions to explain the malfunction.
\texttt{BugDoc} identifies the ground truth
malfunction in $50\%$ of the runs when allowed to run fewer than $10$ interventions. It  identifies the mismatch with intervention budget of $20$ but the returned
solution of \pvt{}s is not minimal. For instance, \texttt{BugDoc} returns
two \pvt{}s: $\{\langle$\pl{$\correlation$}, $\texttt{target}$, $\texttt{education}$$\rangle$
\texttt{and}    $\langle$\pl{$\correlation$}, $\texttt{target}$, $\texttt{sex}$$\rangle\}$
as the explanation of malfunction.

\smallskip

\paragraph{Cardiovascular Disease Prediction.} 
\looseness-1
This system
trains an AdaBoost classifier~\cite{adaboost} on patients' medical
records~\cite{cardio} containing age, height (in centimeters) and weight along with
other attributes. It predicts if the patient
has a disease and does not optimize for false positives. Therefore, the system calculates recall over the patients having cardiovascular
disease, and the goal is to achieve more than $0.70$ recall. The pipeline returns
the additive inverse of recall as the malfunction score. 
We tested the pipeline with two datasets generated through a random selection of records: (1)~the passing dataset 
satisfies the format assumptions of the
pipeline;  (2)~for the failing dataset we manually converted \texttt{height} to inches. 
\systemgr identifies $86$ discriminative \pvt{}s with \texttt{height},
\texttt{weight} and \texttt{age} having the highest degree of $15$ in the \pvt-attribute
graph. Among the \pvt{}s involving these attributes, the \domain of \texttt{height}
has the maximum benefit, which is the ground-truth \pvt too. \systemgr alters the failing dataset by applying a linear transformation
 and it reduces the malfunction from $0.71$ to $0.30$. This explanation matches
the ground truth difference in the passing and the failing dataset. 
Among baselines, \texttt{BugDoc} and \texttt{Anchor} performed $100$ and $5900$
interventions, respectively. Group testing techniques are not applicable
because assumption A3 (Section~\ref{sec:grouptest}) does not hold. We observe
that the malfunction score with a composition of transformation functions is higher than
the one in the original dataset if the composition involves $\correlation$ \pvt. This behavior is observed
because adding noise to intervene with respect to $\correlation$ PVT worsens the classifier performance.
If we remove \pvt{}s that violate this assumption, then \systemgp and \texttt{GrpTest} require $6$ and $9$
interventions, respectively.
 
\smallskip 

\paragraph{Efficiency.} Figure~\ref{tab:comparison} presents 
the execution time 
of considered techniques 
for real-world applications presented above. 
\systemgr, \systemgp and \texttt{GrpTest} are highly efficient and
require less than $30$ seconds to explain the ground-truth cause of
malfunction. In contrast, \texttt{Anchor} is extremely inefficient as it
requires more than 143 minutes for \texttt{cardiovascular}, while \systemgr and
\texttt{BugDoc} explain the malfunction within 63 seconds.

\smallskip

\paragraph{Key takeaways.} Among all real-world case studies, the greedy
approach \systemgr requires the fewest interventions to explain the cause of
malfunction. Group testing techniques, \systemgp and \texttt{GrpTest}, require
fewer interventions than \texttt{BugDoc} and \texttt{Anchor} whenever
assumption A3 (Section~\ref{sec:grouptest}) holds. \texttt{Anchor} requires the
highest number of interventions as it performs many local transformations to
explain the cause of failure. \texttt{BugDoc} optimizes interventions by
leveraging combinatorial design algorithms: it requires more interventions than
\system but fewer than \texttt{Anchor}.

\subsection{Synthetic Pipelines} We evaluate the effectiveness and scalability 
of \systemgr and \systemgp for a diverse set of synthetic scenarios.

\smallskip
\paragraph{\systemgr vs. \systemgp.} In this experiment, we consider a pipeline
where the ground-truth explanation of malfunction violates the observations
discussed in Section~\ref{sec:observations}. Specifically, the
explanation requires modifying one particular value in the dataset and its
likelihood (as estimated by \systemgr) is ranked $54$ among the set of
discriminative \pvt{}s. Therefore, it requires $54$ 
interventions to explain the cause of
malfunction. On the other hand, \systemgp performs group interventions and
requires only $9$ interventions. This experiment demonstrates that \systemgp
requires fewer interventions than \systemgr when the failing dataset and the
corresponding \pvt{}s do not satisfy the observations \systemgr relies on. 
We present additional experiments with complex conjunctive and disjunctive explanations of 
malfunction in the appendix.

\smallskip

\begin{figure}[t]
    \centering
    \includegraphics[width=1\columnwidth]{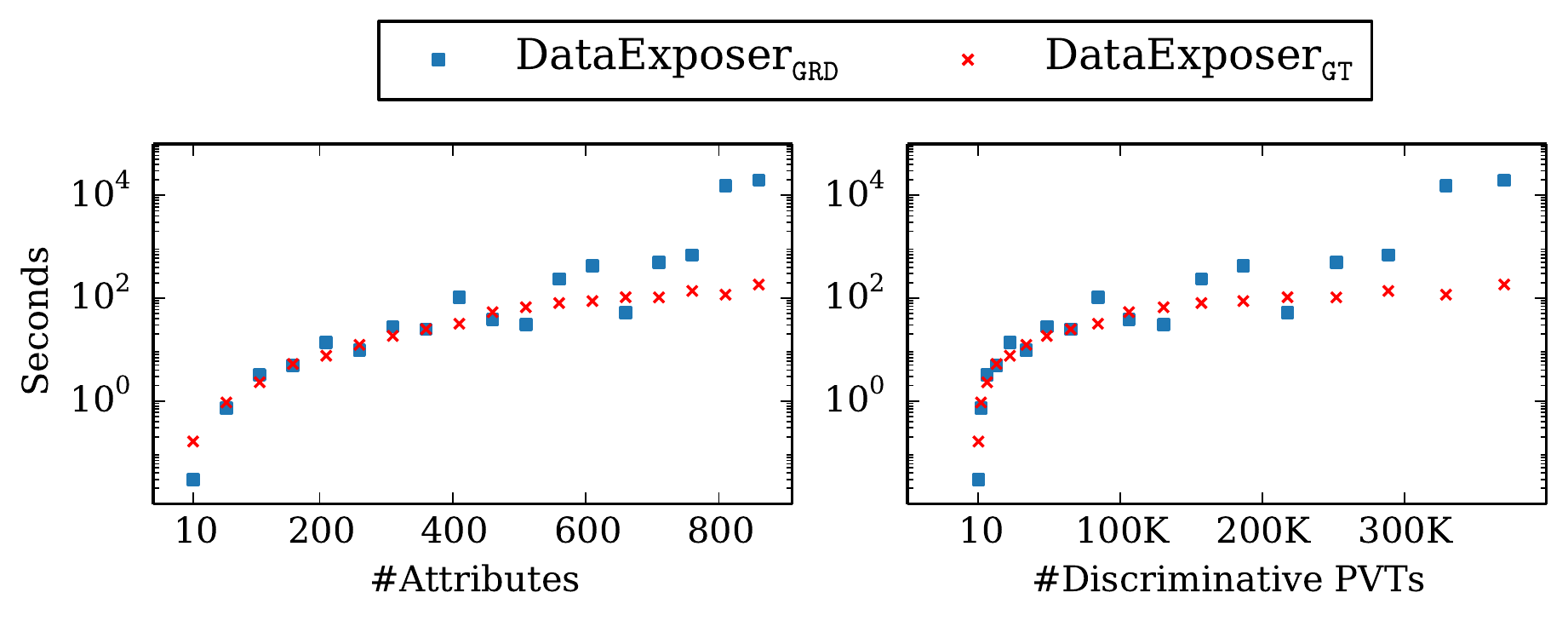}
    \vspace{-5mm}
	
	\caption{Execution time of \systemgr vs. \systemgp with varying number of data attributes (left) and discriminative \pvt{}s (right) over synthetic pipelines.}
	
    \label{fig:scalability}
    \vspace{-1mm}
\end{figure}

\paragraph{Scalability.} To test the scalability of our techniques, we compare
their running time with increasing number of attributes and discriminative
\pvt{}s. Figure~\ref{fig:scalability} shows that the time required by \systemgr
and \systemgp to explain the malfunction grows sub-linearly in the number of
attributes and discriminative \pvt{}s. We observe similar trend of the number
of required interventions on varying these parameters. This experiment
demonstrates that \systemgr requires fewer than $O(|\P|)$ interventions in
practice (where $\P$ denotes the set of discriminative profiles) and validates
the logarithmic dependence of \systemgp on $|\P|$.

\section{Related Work}\label{sec:related}
%

\paragraph{Interventional debugging.} \looseness-1 AID~\cite{aid} uses an
interventional approach to blame runtime conditions of a program for causing
failure; but it is limited to software bugs and does not intervene on datasets.
BugDoc~\cite{bugdoc} finds parameter settings in a black-box pipeline as root
causes of pipeline failure; but it only reports whether a dataset is a root
cause and does not explain why a dataset causes the failure.
CADET~\cite{iqbalcadet} uses causal inference to derive root causes of
non-functional faults for hardware platforms focused on performance issues.
Capuchin~\cite{DBLP:conf/sigmod/SalimiRHS19} casts fairness in machine learning
as a database repair problem and adds or removes rows in the training data to
simulate a \emph{fair world}; but it does not aim to find cause of unfairness.

\smallskip

\paragraph{Data explanation.} \looseness-1 Explanations for query results have
been abundantly studied~\cite{DBLP:conf/sigmod/WangDM15,
Bailis:2017:MPA:3035918.3035928, Chirigati:2016:DPM:2882903.2915245,
GebalyFGKS14}. Some works find causes of errors in data generation
processes~\cite{DBLP:conf/sigmod/WangDM15}, while others discover relationships
among attributes~\cite{Bailis:2017:MPA:3035918.3035928,GebalyFGKS14}, and
across datasets~\cite{Chirigati:2016:DPM:2882903.2915245}.
ExceLint~\cite{barowy2018excelint} exploits the spatial structure of
spreadsheets to look for erroneous formulas. Unlikely interventional efforts,
these approaches operate on observational data, and do not generate additional
test cases.

\smallskip

\paragraph{Model explanation.} \looseness-1 Machine learning
interpreters~\cite{lime,anchors} perturb testing data to learn a surrogate for
models. Their goal is not to find mismatch between data and models. Debugging
methods for ML pipelines are similar to data
explanation~\cite{cadamuro2016debugging, breck2019data}, where training data
may cause model's underperformance.~\cite{varma2017flipper}
and~\cite{kulesza2015principles} discuss principled ways to find reasons of
malfunctions. Wu et al.~\citep{wu2020complaint} allow users to \emph{complain}
about outputs of SQL queries, and presents data points whose removal resolves
the complaints.~\cite{DBLP:conf/sigmod/SchelterRB20} validates when models fail
on certain datasets and assumes knowledge of the mechanism that corrupts the
data. We aim to find discriminative profiles among datasets without such
knowledge.

\smallskip

\begin{sloppypar} \paragraph{Causal inference debugging.} \looseness-1
Data-driven approaches have been taken for causal-inference-based fault
localization~\cite{Gulzar2018,Attariyan2011, Attariyan2012, chen2002pinpoint},
software testing~\cite{fraser@tse2013, godefroid@ndss2008, holler@uss2012,
DBLP:conf/icse/JohnsonB20, DBLP:conf/esec/Zeller99}, and statistical
debugging~\cite{Zheng2006, liblit2005scalable}. However, they use a white-box
strategy or are application-specific. Causal relational
learning~\cite{DBLP:conf/sigmod/SalimiPKGRS20} infers causal relationships in
relational data, but it does not seek mismatches between the data and the
systems. Our work shares similarity with
BugEx~\cite{DBLP:conf/issta/RobetalerFZO12}, which generates test cases to
isolate root causes. However, it assumes complete knowledge of the program, and
data-flow paths. \end{sloppypar}

\smallskip

\paragraph{Data debugging.} \looseness-1 Porting concepts of debugging from
software to data has gained attention in data management
community~\cite{mucslu2013data,brachmann2019data}.
Dagger~\cite{DBLP:conf/cidr/RezigCSSMTOS20, DBLP:journals/pvldb/RezigBTO0MMS20}
provides data debugging primitives for white-box interactions with data-driven
pipelines. CheckCell~\cite{DBLP:conf/oopsla/BarowyGB14} ranks data cells that
unusually affect output of a given target. However, it is not meant for large
datasets where single cells are unlikely to causes malfunction. Moreover,
CheckCell cannot expose combination of root causes. \system is general-purpose,
application-agnostic, and interventional, providing causally verified issues
mismatch between the data and the systems.

\section{Summary and Future Directions}\label{sec:summary}
%

We introduced the problem of identifying causes and fixes of mismatch between
data and systems that operate on data. To this end, we presented \system, a
framework that reports violation of data profiles as causally verified root
causes of system malfunction and reports fixes in the form of transformation
functions. We demonstrated the effectiveness and efficacy of \system in
explaining the reason of mismatch in several real-world and synthetic
data-driven pipelines, significantly outperforming the state of the art. In
future, we want to extend \system to support more complex classes of data
profiles. Additionally, we plan to investigate ways that can facilitate
automatic repair of both data and the system guided by the identified data
issues.

\bibliographystyle{ACM-Reference-Format}
\bibliography{paper}

\ifTechRep
\appendix

%

\section{Minimum Bisection} Algorithm~\ref{alg:minbisection} presents the
\textit{local search algorithm}~\cite{fellows2012local} that \systemgp uses to
partition the set of discriminative \pvt{}s. It is an anytime
algorithm~\cite{zilberstein1996using} and is known to be very efficient in most
practical settings~\cite{fellows2012local}.

\begin{description}

\item[\algoexp{Line~\ref{bs:line1}}] Initialize the solution $\P_1,\P_2$ to a
randomized partition of $\P$ where $|\P_1| - |\P_2| \leq 1$.

\item[\algoexp{Lines~2-\ref{bs:line13}}] Iteratively try to improve the
solution by performing local search.

\item[\algoexp{Line~\ref{bs:line3}}] Count the number of cut edges, i.e.,
edges that connect two nodes from different partitions.

\item[\algoexp{Line~\ref{bs:line3p}}] Initialize the number of cut edges in an
improved partitioning to be equal to the current number of cut edges.

\item[\algoexp{Lines~\ref{bs:line4}-\ref{bs:line7}}] Swap a pair of \pvt{}s
between the partitions to create two temporary partitions.

\item[\algoexp{Lines~\ref{bs:line8}}] Count the number of cut edges between
the temporary partitions.

\item[\algoexp{Lines~\ref{bs:line9}-\ref{bs:line12}}] If swapping the \pvt
pairs reduces the number of cut edges, then update the partitions and terminate
the current iteration.

\item[\algoexp{Line~\ref{bs:line13}}] Stop searching for an improved
partitioning reduction in the number of cut edges is no longer possible.

\item[\algoexp{Line~\ref{bs:line14}}] Return the updated partitions $\P_1,\P_2$
as the solution.

\end{description}

\section{Modeling Interaction among \pvt Interventions}\label{multiple}

We now extend our techniques for the setting when intervening on one \pvt may
affect the impact of intervention on another \pvt. For example, when altering
the failing dataset with respect to \pvt $P_1$ does not help reduce malfunction
when considered in isolation, but, malfunction reduces when the failing dataset
is intervened with respect to \pvt $P_1$ along with another \pvt $P_2$. In such
a setting, assumption A2 does not hold and \system may fail to identify the
ground-truth cause of malfunction.

To address this issue, we propose to leverage multiple passing and failing
datasets. Using this additional knowledge, we generate a decision tree to guide
\system. The key idea is to fit a decision tree based on the \pvt{}s and the
relative decrease of the malfunction score ($m_S{D}$) with respect to the
maximum allowable malfunction score threshold ($\tau$). Each training data
point for learning the decision tree consists of the set of \pvt{}s $\P_i$,
extracted from dataset $D_i$, and the outcome is the Boolean assessment of
whether $m_S(D_i) \le \tau$. Each path of the tree can be seen as a conjunction
of \pvt{}s. If a conjunction leads to a consistent outcome (a ``pure'' leaf),
then that combination becomes an explanation ``candidate''. The candidates are
then verified by Algorithm~\ref{alg:datadebugthree}, where we perform
interventions on an arbitrary failing dataset.

\begin{description}
\item[\algoexp{Line~\ref{dt:init}}] Traverse the decision tree at least once.

\item[\algoexp{Line~\ref{dt:while}-\ref{dt:line11}}] Follow the decision-tree
paths until we find a conjunction of \pvt{}s $\P$ that reduces the malfunction
score of $D_{\mathit{fail}}$ so that $m_S(D_{\mathit{fail}}) \le \tau$. If no
candidate conjunction of \pvt{}s can convert $D_{\mathit{fail}}$ to a passing
dataset, then rebuild the decision tree.

\begin{algorithm}[t]
		\LinesNumbered
		\DontPrintSemicolon
\KwIn{Dependency Graph $G_{PD}$, discriminative \pvt{}s $\P$}
\KwOut{Partitions $\P_1$, $\P_2$}
$\P_1,\P_2\leftarrow \textsc{Random-Bisection}(\P)$\label{bs:line1} \tcc*[f]{Initialization}\\
\Do{$C' < C$;\tcc*[f]{As long as cut size reduces}\label{bs:line13}}
	{
		$C \leftarrow  \sum\limits_{X_1\in \P_1, X_2\in \P_2} \mathbbm{1}_{G _{PD}}(X_1,X_2 )$\label{bs:line3}\tcc*[f]{\#Cut edges}\\
		$C' \leftarrow C$\label{bs:line3p}\tcc*[f]{Initialize}\\
		\ForEach{$X_1 \in \P_1$\label{bs:line4}}{
		 \ForEach{$X_2 \in \P_2$\label{bs:line5}}{
			$\P_1'\leftarrow \P_1\cup\{X_2\}\setminus \{X_1\}$\label{bs:line6}\tcc*[f]{Swap \pvt{}s }\\ 
			$\P_2'\leftarrow \P_2\cup\{X_1\}\setminus \{X_2\}$\label{bs:line7}\tcc*[f]{Swap \pvt{}s }\\
			$ C' \leftarrow  \sum\limits_{X_1\in \P_1', X_2\in \P_2'} \mathbbm{1}_{G _{PD}}(X_1,X_2 )$\label{bs:line8}\tcc*[f]{\#Cut edges}\\
			\If(\tcc*[f]{Cut size is reduced}){$C' < C$\label{bs:line9}}
			{
				$\P_1 \leftarrow \P_1'$ \label{bs:line10}\tcc*[f]{Update partition}\\
				$\P_2 \leftarrow \P_2'$ \label{bs:line11}\tcc*[f]{Update partition}\\
				\textbf{break}\label{bs:line12}
			}
		 }
	  }
   }
\Return $\P_1,\P_2$ \label{bs:line14}\tcc*[f]{$\P_1,\P_2$ are returned as the solution}
\caption{\textsc{Get-Min-Bisection}}
\label{alg:minbisection} 
\end{algorithm}

\item[\algoexp{Line~\ref{dt:lineone}}] Find the paths $\Pi$ of the decision
tree that lead to pure leaves (malfunction score is less than $\tau$ whenever
all \pvt{}s in the paths, in conjunction, are observed) and sort them by their
\pvt{}s' benefit scores, as defined in Algorithm~\ref{alg:datadebuggreedy}.

\item[\algoexp{Line~\ref{dt:unmark}}] Record that some candidate \pvt{}s were
computed.

\item[\algoexp{Lines~\ref{dt:linetwo}-\ref{dt:line11}}] Iteratively test each
conjunction of \pvt{}s $\P$.

\item[\algoexp{Line~\ref{dt:linethree}}] Transform the failing dataset
$D_{\mathit{fail}}$ w.r.t $\P$ to obtain $D_t$.

\item[\algoexp{Lines~\ref{dt:lineeleven}-\ref{dt:line11}}] If the transformed
dataset $D_t$ reduces the malfunction score, then return $\P$ as a cause of
malfunction. Otherwise, add a new instance (data point) to re-train the
decision tree and recompute $\Pi$.

\end{description}

In future work, we plan to extend this approach to leverage
combinatorial-design-based techniques to efficiently intervene \pvt{}s to
initialize the decision tree algorithm~\cite{JCD:JCD20065}.

\section{Discussion on \systemgr vs. \systemgp} As discussed in
Section~\ref{sec:grouptest}, \systemgr always identifies the ground-truth cause
of malfunction, but, the number of interventions required may increase if
observations O1-O3 do not hold. If O1 does not hold, then the initial ordering
of attributes, which is computed based on their degree in \pvt{}-Attribute
graph $G_{PA}$, may not be accurate. In this case, \systemgr may require
$O(\P)$ interventions to explain the cause of system malfunction. Similarly,
whenever O2 or O3 fail, the benefit-based ordering of the \pvt{}s is likely to
be incorrect. In applications where O1 holds but O2 and O3 do not hold,
$G_{PA}$ is expected to be accurate and the number of required interventions is
$O(r)$, where $r$ is the degree of the attribute that has the highest degree in
$G_{PA}$. In contrast, \systemgp requires $O(t\log |\P|)$ interventions even
when these observations are violated. Therefore, \systemgr is beneficial
whenever all of these observations hold, or $t=\Omega(|\P|/\log |\P|)$.

\begin{algorithm}[t]
	\LinesNumbered
\KwIn{Failing dataset $D_{\mathit{fail}}$, decision tree $DT$, malfunction score threshold $\tau$}
$\textsc{Update-Paths} \leftarrow \texttt{true}$\label{dt:init}\tcc*[f]{Indicates whether to explore decision tree paths}\\
\While{$\textsc{Update-Paths}$\label{dt:while}}{
    Let $\Pi$ be the set of  paths in $DT$ that are observed for instances where system malfunction score is less than $\tau$. All \pvt{}s along a path are considered in conjunction and sorted  by decreasing order of their benefit score.\label{dt:lineone} \\
    
    $\textsc{Update-Paths} \leftarrow \texttt{false}$\label{dt:unmark}\tcc*[f]{$\Pi$ is up-to-date}\\
    
    \For{$\P\in \Pi$\label{dt:linetwo}}
    {
    	$D_t \leftarrow \P_\T(D_{\mathit{fail}})$\label{dt:linethree}
    	\tcc*[f]{Transform $D_{\mathit{fail}}$ wrt $\P$}\\
    	
    	\If(\tcc*[f]{Cause has been identified}){$m_S(D_t) \le \tau$\label{dt:lineeleven}}
    	{
    		\Return $\P$\label{dt:linetwelve}
    	}
    	\Else{
    	$DT.\textsc{Update}(D_t, \texttt{Fail})$\label{dt:linefive}
    	\tcc*[f]{Add instance to tree}\\
    	
    	    $\textsc{Update-Paths} \leftarrow \texttt{true}$\label{dt:line11}\tcc*[f]{Recompute $\Pi$}\\
    	}
    }
}
\Return $\emptyset$
\caption{Greedy + Decision Tree Approach}
\label{alg:datadebugthree} 
\end{algorithm}

\section{Effect of Various Parameters}

In this experiment, we test the effect of the number of dataset attributes and
the number of discriminative \pvt{}s on the efficacy of \system algorithms, and
contrast those with other state-of-the-art baselines. We also investigate the
influence of the number of \pvt{}s involved in the root causes and their
interactions on the number of interventions each method requires.

\begin{figure*}[t]
    \centering
    \includegraphics[width=\textwidth]{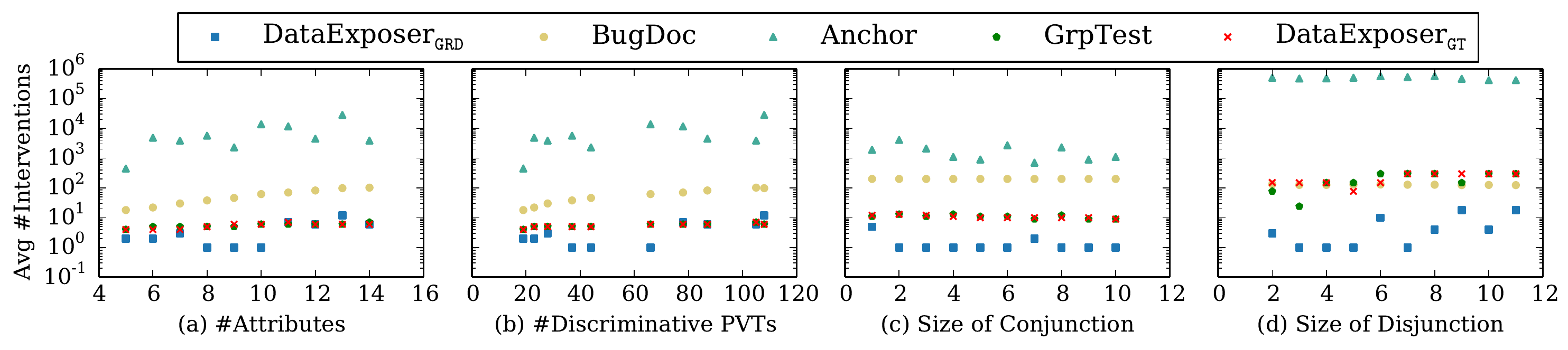}
	
	 \caption{Average number of interventions required by two versions of
	 \system and three other techniques for varying number of attributes,
	 discriminative \pvt{}s, size of single conjunctive root causes, and size of
	 disjunctive root causes.}
    
	\label{fig:interventions}
\end{figure*}

\subsection{Effect of Number of Attributes and \pvt{}s}
Figure~\ref{fig:interventions}(a)~presents the effect of changing the number of
attributes in the datasets on the number of required interventions.
\systemgr requires fewer than $5$ interventions on average. In contrast,
\texttt{BugDoc} and \texttt{Anchor} require orders of magnitude more
interventions. The number of interventions required by \texttt{BugDoc} grows
linearly with the number of attributes. At the same time, \texttt{Anchor}
perturbs all \pvt{}s to solve a multi-armed bandit problem: the more \pvt{}s
affect the pipeline errors, the more interventions are needed.
Group-testing-based approaches also require more intervention than that of
\systemgr, and grow logarithmically with the number of data attributes.

Figure~\ref{fig:interventions}(b)~depicts the effect of the number of
discriminative \pvt{}s on the number of required interventions. \systemgr shows
superior performance as it requires fewer than $10$ interventions even when the
number of discriminative \pvt{}s go beyond $100$. Here, we observe trends
similar the one in Figure~\ref{fig:interventions}(a) for other baselines as the
number of \pvt{}s are strongly and positively correlated to the number
attributes.

\balance

\subsection{Effect of Number of Root Causes and their Interactions} The
pipelines presented in Figures~\ref{fig:interventions}(a)
and~\ref{fig:interventions}(b) have a single \pvt as the root cause of the
malfunction. In Figure~\ref{fig:interventions}(c), we fix the number of
attributes to $15$ and the number of discriminative \pvt{}s between the passing
and the failing datasets to $136$. We also define the root cause to be a
conjunction over a set of \pvt{}s of varying cardinalities. We find that the
cardinality of the root-cause set (length of the conjunctive cause) does not
impact the number of interventions as much as the number of attributes and the
number of discriminative \pvt{}s do. However, having more than one cause for
malfunction (i.e., a disjunctive cause) requires more interventions for
\texttt{Anchor} and the Group-Test techniques, as shown in
Figure~\ref{fig:interventions}(d). \systemgr still needs orders of magnitude
fewer interventions than these other approaches, although the probability of
failing to find any feasible transformation, which decreases malfunctions
scores, increases with the number of possible root causes within the
disjunction.

\fi 

\end{document}